\documentclass[preprint]{imsart}
\usepackage{amsmath}
\usepackage[pdftex]{graphicx}
\usepackage{float}
\usepackage{caption}
\usepackage[hyphens,spaces]{url}
\usepackage{amsthm}
\usepackage{bbm}
\usepackage{array,multirow}
\usepackage{pbox}
\usepackage{algorithm, algorithmic,refcount}
\usepackage{color}
\usepackage{epstopdf}
\usepackage{tikz}
\usetikzlibrary{positioning}
\usetikzlibrary{arrows,automata}
\tikzset{
  supset/.style={
    draw=none,
    edge node={node [sloped,  auto=false]{$\supset$}}},
  Subset/.style={
    draw=none,
    every to/.append style={
      edge node={node [sloped, allow upside down, auto=false]{$\subset$}}}
  }
}

\DeclareMathAlphabet      {\mathbfit}{OML}{cmm}{b}{it}

\usepackage{pgfplots}
\usepgfplotslibrary{units}
\usepackage{subcaption}

\newtheorem{result}{Result}
\usepackage{multirow}% http://ctan.org/pkg/multirow
\usepackage{amssymb}
\usepackage{times,parskip,epsf,epsfig,latexsym,rotating,natbib}
\usepackage{graphicx}

\begin{document}

\begin{frontmatter}
\title{Manifold valued data analysis of samples of networks, with applications in corpus linguistics\thanksref{T1}}
\runtitle{Manifold valued data analysis of samples of networks} 

\thankstext{T1}{This work was supported by the Engineering and Physical Sciences Research Council [grant number EP/T003928/1 and 
EP/M02315X/1]. The authors are grateful 
to Michaela Mahlberg, Viola Wiegand and Anthony Hennessey for their help and discussions about the data, and to the Editor, Associate Editor and two anonymous referees for their very helpful comments.}

\begin{aug}
  \author{\fnms{Katie E.}  \snm{Severn},
  %\corref{}
  %\thanksref{t2}
  \ead[label=e1]{katie.severn@nottingham.ac.uk}}
  \author{\fnms{Ian L.} \snm{Dryden}\ead[label=e2]{ian.dryden@nottingham.ac.uk}}
  \and
  \author{\fnms{Simon P.}  \snm{Preston}%
  \ead[label=e3]{simon.preston@nottingham.ac.uk}%
  \ead[label=u1,url]{http://www.foo.com}}

%\thankstext{t2}{text for thanks}

  \runauthor{K.E.SEVERN, I.L.DRYDEN and S.P.PRESTON}

  \affiliation{University of Nottingham}

  \address{K.E.SEVERN, I.L.DRYDEN and S.P.PRESTON,\\ 
  SCHOOL OF MATHEMATICAL SCIENCES\\
UNIVERSITY OF NOTTINGHAM\\
UNIVERSITY PARK\\
NOTTINGHAM, NG7 2RD\\
UK\\
%E-MAIL: ian.dryden@nottingham
          \printead{e1,e2,e3}}

  %\address{Address of the Third author,\\
   %       \printead{e3,u1}}

\end{aug}
\renewcommand{\thesubfigure}{\roman{subfigure}}

\makeatletter
\newcommand\approxsim{\mathchoice
  {\@approxsim {\displaystyle}      {1ex} }
  {\@approxsim {\textstyle}         {1ex} }
  {\@approxsim {\scriptstyle}       {.7ex}}
  {\@approxsim {\scriptscriptstyle} {.5ex}}}
\newcommand\@approxsim[2]{%
  \mathrel{%
    \ooalign{%
      $\m@th#1\sim$\cr
      \hidewidth$\m@th#1.$\hidewidth\cr
      \hidewidth\raise #2 \hbox{$\m@th#1.$}\hidewidth\cr
    }%
  }%
}
\makeatother

\begin{abstract}
Networks arise in many applications, such as in the analysis of text documents, social interactions and brain activity. We develop a general framework for extrinsic statistical analysis of samples of networks, motivated by networks representing text documents in corpus linguistics. We identify networks with their graph Laplacian matrices, for which we define metrics, embeddings, tangent spaces, and a projection from Euclidean space to the space of graph Laplacians.  This framework provides a way of computing means, performing principal component analysis and regression, and carrying out hypothesis tests, such as for testing  for equality of means between two samples of networks. We apply the methodology to the set of novels by Jane Austen and Charles Dickens.
\end{abstract}



\end{frontmatter}

\section{Introduction}
\makeatletter
\newcommand\approxsim{\mathchoice
  {\@approxsim {\displaystyle}      {1ex} }
  {\@approxsim {\textstyle}         {1ex} }
  {\@approxsim {\scriptstyle}       {.7ex}}
  {\@approxsim {\scriptscriptstyle} {.5ex}}}
\newcommand\@approxsim[2]{%
  \mathrel{%
    \ooalign{%
      $\m@th#1\sim$\cr
      \hidewidth$\m@th#1.$\hidewidth\cr
      \hidewidth\raise #2 \hbox{$\m@th#1.$}\hidewidth\cr
    }%
  }%
}
\makeatother

The statistical analysis of networks dates back to at least the 1930's, however interest has increased considerably
in the 21st century \citep{kolaczyk2009statistical}.  Networks are able to represent many different types of data,
for example social networks, neuroimaging data and text documents.  In this paper, each observation is
a weighted network, denoted $G_m=(V,E)$, comprising a set of nodes, $V=\lbrace v_1, v_2,\ldots, v_m\rbrace$, and
a set of edge weights, $E=\lbrace w_{ij} : w_{ij}\geq 0, 1\leq i,j \leq m\rbrace$, indicating nodes $v_i$ and
$v_j$ are either connected by an edge of weight $w_{ij}>0$, or else unconnected (if $w_{ij}=0$). An unweighted
network is the special case with $w_{ij}\in\lbrace 0,1\rbrace$.  We restrict attention to networks that are
undirected and without loops, so that $w_{ij}=w_{ji}$ and $w_{ii}=0$, then any such network can be identified with
its graph Laplacian matrix $\textbf{L}=(l_{ij})$, defined as
\begin{align*}
   l_{ij} = 
\begin{cases}
    -w_{ij}, & \text{if } i\neq j\\
    \sum_{k\neq i}w_{ik},& \text{if } i=j
\end{cases}
\end{align*}  
for $1\leq i,j \leq m$. 

The graph Laplacian matrix can be written as $\textbf{L}=\textbf{D}-\textbf{A}$, in terms of the
adjacency matrix, $\textbf{A}=(w_{ij})$, and degree matrix 
$\textbf{D}=\text{diag}(\sum_{j=1}^mw_{1j},\ldots,\sum_{j=1}^mw_{mj})=\text{diag}(\textbf{A}\textbf{1}_m)$, where
$\textbf{1}_m$ is the $m$-vector of ones.  The $i$th diagonal element of $\textbf{D}$ equals the degree of node
$i$.   The space of $m \times m$ graph Laplacian matrices is 
\begin{align}\label{eq:lapl space}
\mathcal{L}_m=\lbrace \textbf{L}=(l_{ij}):\textbf{L}=\textbf{L}^T ;\, l_{ij}\leq 0 \, \forall i\neq j ;\, \textbf{L} {\textbf{1}_m}={\textbf{0}_m} \rbrace,
\end{align}
where $\textbf{0}_m$ is the $m$-vector of zeroes. The space $\mathcal{L}_m$ is a closed convex subset of the cone of centred 
symmetric 
positive semi-definite $m\times m$ matrices:
\begin{align}
\mathcal{PSD}^*_m=\lbrace \textbf{S}^{m\times m} : x^T\textbf{S}x\geq 0 \, \forall x \in \mathbb{R}^m; \, \textbf{S}=\textbf{S}^T; \textbf{S} {\textbf{1}_m}={\textbf{0}_m} \rbrace , 
\end{align}
and $\mathcal{L}_m$ is a manifold with corners \citep{ginestet2017hypothesis}. 
%Also $\mathcal{PSD}^*_m$ is a subset of the more commonly used space of symmetric positive semi-definite matrices (see, for example, \citet{10.2307/30242879} for statistical analysis in this space).  
The relationship ${\mathcal L}_m \subset \mathcal{PSD}^*_m$  is evident as $\textbf{L}\in\mathcal{L}_m$ satisfies $\textbf{L}=\textbf{L}^T$ and $\textbf{L} {\textbf{1}_m}={\textbf{0}_m}$ due to the definition of $\mathcal{L}_m$ in (\ref{eq:lapl space}) and any $\textbf{L}\in\mathcal{L}_m$ is diagonally dominant, as $\vert l_{ii}\vert=\sum_{i\neq j} \vert l_{ij} \vert$,  which is a sufficient condition for any $\textbf{L}\in \mathcal{L}_m$ to satisfy $x^T\textbf{L}x\geq 0 \, \forall x \in \mathbb{R}^m$  \citep[page 232]{de2006aspects}. Both ${\mathcal L}_m$  and $\mathcal{PSD}^*_m$ have dimension $m(m-1)/2$.

For the tasks we address the data are a random sample $\textbf{L}_1,\ldots, \textbf{L}_n$ from a population of networks, 
where each observation is a graph Laplacian $\textbf{L}_k\in\mathcal{L}_m, \; k=1,\ldots,n$ representing networks with a common 
node set $V$. 
Graph Laplacians are not standard Euclidean data and so for typical
statistical tasks such as computing the mean, performing principal component analysis, regression, and two sample tests on means, standard Euclidean methods need to be carefully adapted.

To perform statistical analysis on the manifold of graph Laplacians we need to define suitable metrics. First of all we introduce the Euclidean distance between matrices $\textbf{X}$ and $\textbf{Y}$, also known as the Frobenius distance
\begin{equation}
 d_E(  \textbf{X} ,\textbf{Y} ) =  \Vert\textbf{X} -\textbf{Y} \Vert = \{ {\rm trace} (\textbf{X}-\textbf{Y})^T(\textbf{X}-\textbf{Y}) \}^{\frac{1}{2}} , \label{Euc}
 \end{equation}
and the Procrustes distance 
\begin{equation}
 d_S(  \textbf{X} ,\textbf{Y} ) =   \inf_{\textbf{R}\in \mathcal{O}(m)} \Vert\textbf{X} -\textbf{Y}\textbf{R} \Vert , \label{Proc} 
\end{equation}
which involves optimizing over an orthogonal matrix 
${\textbf{R}}$ for the ordinary 
Procrustes match of $\textbf{Y}$ to $\textbf{X}$ \citep[chapter 7]{MR3559734}. When the matrices are centred, as will be the case throughout the paper, 
this Procrustes distance is also known as the Procrustes size-and-shape distance \citep[chapter 5]{MR3559734}. 

We will consider two general metrics between graph Laplacians in ${\mathcal L}_m$, which are based on these matrix distances. 
The Euclidean power metric between graph Laplacians is  
\begin{equation}
d_\alpha(\textbf{L}_1, \textbf{L}_2) =   d_E( \textbf{L}_1^\alpha,\textbf{L}_2^\alpha ) , \label{eq:euc power metric}
\end{equation} 
and the Procrustes power metric between graph Laplacians is 
\begin{equation}
  d_{\alpha,S}(\textbf{L}_1, \textbf{L}_2) =  d_S( \textbf{L}_1^\alpha,\textbf{L}_2^\alpha )  , \label{eq:proc power metric}
\end{equation}
where the power of the graph Laplacian $\textbf{L}_j^\alpha, j=1,2$ is defined in (\ref{Fmap}). 
Common choices of Euclidean power metrics and Procrustes metrics are $d_1$, $d_\frac{1}{2}$ and $d_{\frac{1}{2}, S}$, referred to as the Euclidean, square root Euclidean and Procrustes size-and-shape metrics respectively \citep{10.2307/30242879}.  We provide more detail about these metrics in Section \ref{sec:space}.

Analysing networks by representing them as elements of $\mathcal{L}_m$ is an approach also used by 
\citet{ginestet2017hypothesis}.  The authors considered the Euclidean metric $d_1$ and derived a central limit 
theorem which
they used to develop a test of mean difference between two samples of networks, driven by an application in neuroimaging.  Motivation
for our consideration of metrics other than $d_1$ 
includes evidence that there can be advantages to using non-Euclidean metrics when interpolating non-Euclidean data, for example less swelling in the context of positive semi-definite matrices \citep{10.2307/30242879}. 
%and improved interpretability (in the context of dynamic networks \citep{bakker2018dynamic}).

\citet{kolaczyk2020} have similarly considered using non-Euclidean metrics for network data. Their `Procrustean distance' for unlabelled networks is different from our Procrustes distance in that they 
restrict their analogue of $\textbf{R}$ in (\ref{Proc}) to be a permutation matrix, whereas we allow it to be a more general orthogonal matrix. 
In addition \citet{kolaczyk2020}  retain symmetry and have $\textbf{R}^T\textbf{Y}\textbf{R}$ rather than $\textbf{Y}\textbf{R}$ in (\ref{Proc}), which we use following \citet{10.2307/30242879}. 
Although the metrics are different, this connection
provides motivation for using our Procrustes metric, for example where nodes need to be relabelled or combined when computing a distance.
Calculating the Procrustes metric is more straightforward when optimising over orthogonal matrices compared to permutations, 
and so orthogonal Procrustes provides a fast approximation for Procrustes matching with permutations.

% The central limit theorem involves estimating a large covariance matrix from a relatively small sample and requires that all off diagonal elements are non zero in expectation. We introduce our own central limit theorem and two sample test in Section %\ref{sec:two sample test}.

%The motivating application for our work is the analysis of text documents.  Networks are used to model documents in a text corpus \citep{phillips1983lexical}. Each node represents a word and edges indicate words co-occurring within a span of, say, 5 words of each other.  %This representation conserves information on the co-occurrence of words, and these co-occurrences can be distinctive for different texts, be it authors or genres. 
%More background on co-occurrence and its use in corpus linguistics can be found in \citet{evert2008corpora}. By representing these texts networks as graph Laplacians we provide a way of answering questions such as ``what is the mean of a set %of texts?'' and ``how can we test if different texts are written by the same author?''. 
%A recent example of testing authorship was for the novel {\it The Cuckoo's Calling} written under the pen name `Robert Galbraith', but later found to be written by the famous J.K. Rowling \citep{juola2015rowling}. 
%The two sets of texts we analyse are Jane Austen and Charles Dickens novels, found on CLiC \citep{doi:10.3366/cor.2016.0102} and explained further in Section \ref{sec:dickens and austen}. 

\section{Application: Jane Austen and Charles Dickens novels}\label{sec:dickens and austen} 
In corpus linguistics, networks are used to model documents comprising a text corpus \citep{phillips1983lexical}. Each node
represents a word, and edges indicate words that co-occur within some span---typically 5 words, which we use
henceforth---of each other 
\citep{evert2008corpora}.  Our dataset is derived from the full text in novels\footnote{ {\it Christmas Carol} and {\it
Lady Susan} are short novellas rather than novels, but we shall use the term ``novel" for each of the works for
ease of explanation. }  by Jane Austen and Charles Dickens, as listed in Table \ref{table:novels}, obtained from
CLiC \citep{doi:10.3366/cor.2016.0102}. For each of the 7 Austen and 16 Dickens novels, the ``year written'' 
refers to the year in which the author started writing the novel; see \citet{janeAus} and \citet{charlesDickens}.  Our key
statistical goals are to investigate the authors' evolving writing styles, by regressing the networks on ``year
written''; to explore dominant modes of variability, by developing principal component analysis for samples of
networks; 
and to test for significance
of differences in Austen's and Dickens' writing styles, via a two-sample test of equality of mean networks.

\begin{table}[t] 
\centering
\begin{tiny}
\begin{tabular}{|c |c| c|c|} 
\hline
\textbf{Author} & \textbf{Novel name} & \textbf{Abbreviation} & \textbf{Year written} \\
\hline 
 Austen & Lady Susan& LS&   1794\\ \hline 
 Austen & Sense and Sensibility& SE&      1795 \\\hline 
 Austen & Pride and Prejudice& PR&    1796\\\hline
 Austen & Northanger Abbey& NO&    1798\\ \hline
 Austen & Mansfield Park& MA&    1811\\ \hline
 Austen & Emma& EM&  1814\\\hline
 Austen & Persuasion & PE&    1815\\\hline
 Dickens  & The Pickwick Papers& PP&1836\\\hline
 Dickens & Oliver Twist& OT&1837  \\ \hline
 Dickens & Nicholas Nickleby & NN&1838\\ \hline 
 Dickens & The Old Curiosity Shop& OCS& 1840\\\hline
 Dickens & Barnaby Rudge&  BR &  1841  \\\hline
 Dickens & Martin Chuzzlewit& MC& 1843\\ \hline
 Dickens & A Christmas Carol & C& 1843 \\\hline
 Dickens & Dombey and Son& DS& 1846\\\hline
 Dickens & David Copperfield& DC& 1849\\\hline
 Dickens & Bleak House&BH & 1852\\\hline
 Dickens & Hard Times& HT&1854 \\\hline
 Dickens & Little Dorrit& LD& 1855\\\hline
 Dickens & A Tale of Two Cities& TTC& 1859   \\\hline
 Dickens & Great Expectations& GE&1860\\\hline
 Dickens & Our Mutual Friend& OMF& 1864\\ \hline
 Dickens & The Mystery of Edwin Drood & ED& 1870 \\\hline
\end{tabular}
\end{tiny}
\caption{The Jane Austen and Charles Dickens novels from the CLiC database \citep{doi:10.3366/cor.2016.0102}}\label{table:novels}
\end{table}
 
For each Austen and Dickens novel we produce a network representing pairwise word co-occurrence.  
If the node set $V$ corresponded to every word in all the novels it would be very large, with
$m=48285$, but a relatively small number of words are used far more than others. The top $m=50$ words cover
$45.6\%$ of the total word frequency, $m=1000$ cover $79.6\%$, and $m=10000$ cover $96.7\%$. We focus on a
truncated set of the $m$ most frequent words for the combined set of all novels of both authors. and the $w_{ij}$'s are the pairwise co-occurrence counts between these
words. Hence the node set $V$ is consistent between all networks with a common labelling of nodes regardless of novel or author. Although the labelling of the nodes is 
fixed in our applications, the Procrustes metric does allow for some relabelling or combining of
words. For example the metric would be useful where equivalent words or spellings are used (e.g. {\it thy} versus {\it your}) 
and more generally where nodes from different novels/authors are not ordered and they need to be relabelled or combined when computing a distance. 

In our analysis we choose $m=1000$ as a sensible trade-off between having very large, very sparse graph
Laplacians versus small graph Laplacians of just the most common words.   For each novel and the truncated node
set, the network produced is converted to a graph Laplacian. A pre-processing step for the novels is to normalise
each graph Laplacian in order to remove the gross effects of different lengths of the novels by dividing each graph
Laplacian by its own trace, resulting in a trace of 1 for each novel.

  As an indication of the broad similarity of the most common words we list the top 25 words in the table in
  Appendix \ref{TAB2}. Of the top 25 words across all novels 22 appear in the most frequent 25 words for the
  Dickens novels and 23 for the Austen novels. The words {\it not}, {\it be}, {\it she} do not appear in Dickens'
  top 25 and the words {\it mr} and {\it said}  do not appear in Austen's top 25.  Some differences in relative
  rank are immediately apparent: {\it her}, {\it she}, {\it not} having higher relative rank in Austen and {\it
  he}, {\it his}, {\it mr}, {\it said} having relatively higher rank in Dickens.

  %, this gives graph Laplacians $L_k=(\textbf{D}-\textbf{A})/\textbf{1}^T\textbf{A1}$. %Doing so removes length effect as the trace represents the total co-occurrences which will be consistent for each novel now. 

\begin{figure}[htbp]
    \centering
 %  \begin{subfigure}[b]{0.45\textwidth}
 %\caption*{}
  %      \includegraphics[trim={0cm 2cm 0cm 0cm}, clip, scale=1]{dendEuc_1000}
  %  \end{subfigure}
  %  \vspace{-5mm}
  %           \begin{subfigure}[b]{0.45\textwidth}
 %\caption*{}
  %      \includegraphics[trim={0cm 2cm 0cm 0cm}, clip, scale=1]{clustEuc_1000}
  %  \end{subfigure}
  %  \vspace{-5mm}
       \begin{subfigure}[b]{0.45\textwidth}
 \caption*{}
        \includegraphics[trim={0cm 2cm 0cm 0cm}, clip, scale=1]{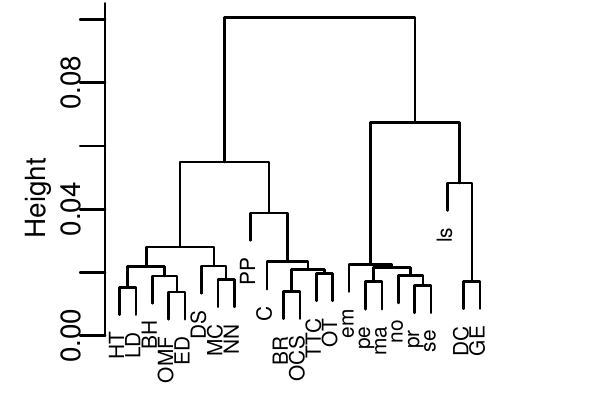}
    \end{subfigure}
             \begin{subfigure}[b]{0.45\textwidth}
 \caption*{}
        \includegraphics[trim={0cm 2cm 0cm 0cm}, clip, scale=1]{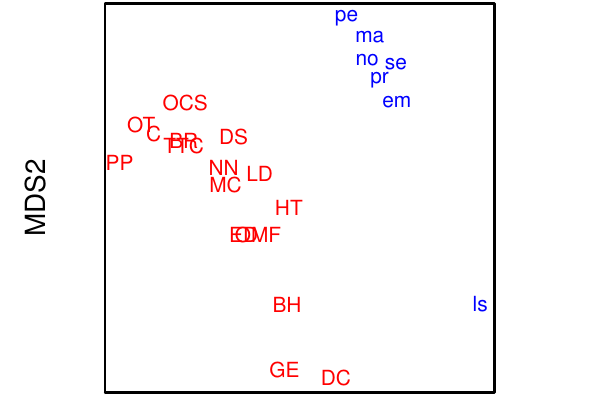}
    \end{subfigure}
    \vspace{-5mm}
       \begin{subfigure}[b]{0.45\textwidth}
 \caption*{}
        \includegraphics[trim={0cm 2cm 0cm 0cm}, clip, scale=1]{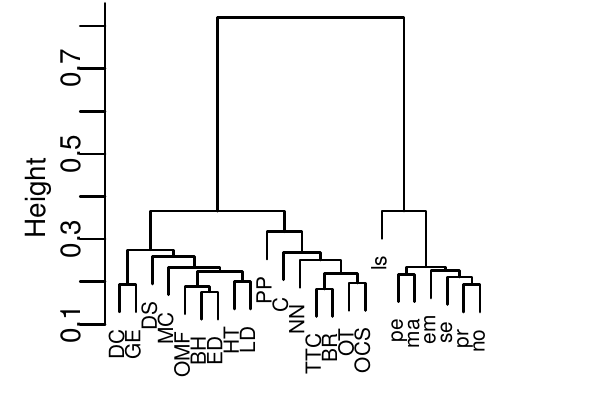}
    \end{subfigure}
             \begin{subfigure}[b]{0.45\textwidth}
 \caption*{}
        \includegraphics[trim={0cm 2cm 0cm 0cm}, clip, scale=1]{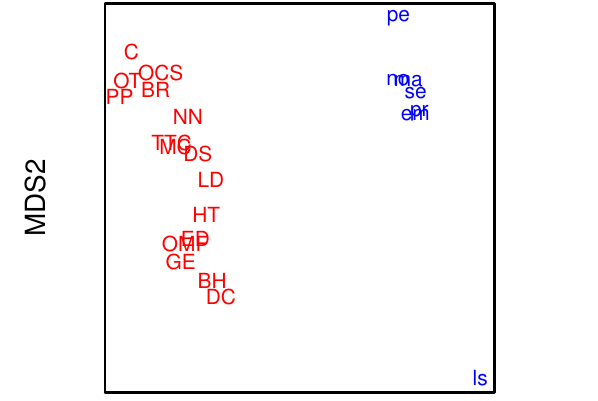}
            \end{subfigure}
    \vspace{-5mm}
           \begin{subfigure}[b]{0.45\textwidth}
 \caption*{}
        \includegraphics[trim={0cm 0cm 0cm 0cm}, clip, scale=1]{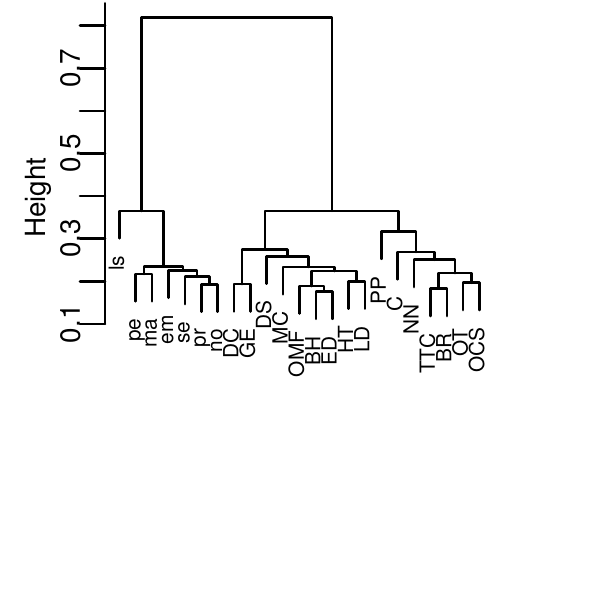}
    \end{subfigure}
             \begin{subfigure}[b]{0.45\textwidth}
 \caption*{}
        \includegraphics[trim={0cm 0cm 0cm 0cm}, clip, scale=1]{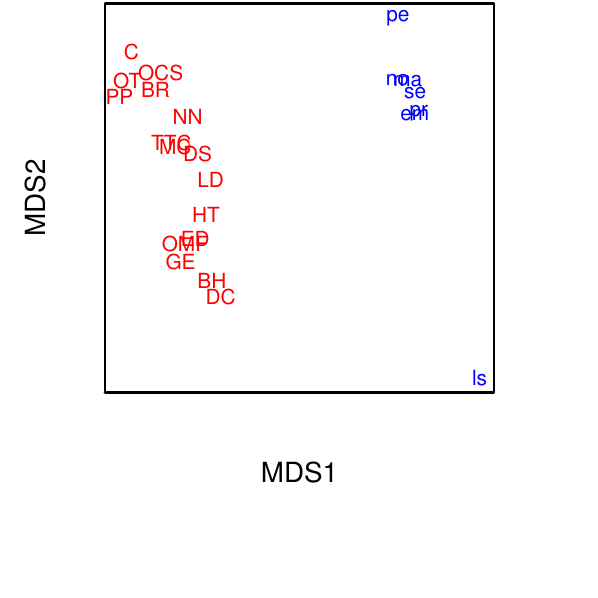}
    \end{subfigure}
  \caption{ \footnotesize{\textit{Cluster analysis and MDS plots based on (from top to bottom) the Euclidean
    distance, $d_1$, square root distance, $d_{\frac{1}{2}}$, and Procrustes distance, $d_{\frac{1}{2}, S}$ each with $m=1000$. The plots display Austen's novels in lower case, and Dickens's novels in upper case.}}}\label{ClusterMDS}
        \end{figure}

We initially compare some choices of distance metrics on the Austen and Dickens data after constructing the graph Laplacians from the $m=1000$ most frequent words across all 23 novels. 
Figure \ref{ClusterMDS} (left column) shows the results of a hierarchical cluster analysis using 
Ward's method \citep{Ward63}, based on pairwise distances between novels using metrics 
$d_1$, $d_\frac{1}{2}$ and $d_{\frac{1}{2}, S}$. For computing the Procrustes metric we use the {\tt shapes} package \citep{Dryden-shapes} in R \citep{CRAN18}. 

The dendrograms for square root and Procrustes 
separate the authors into two
very distinct clusters, whereas for Euclidean distance Dickens' {\it David Copperfield} and {\it Great
Expectations} are clustered with Austen's {\it Lady Susan} which is unsatisfactory. The next sub-division of the
Dickens cluster using square root/Procrustes distance splits into groups of the earlier novels versus later novels,
with the exception being the historical novel {\it A Tale of Two Cities} which is clustered with the earlier
novels. There is not such a clear sub-division for Dickens using the Euclidean metric.  In the Austen cluster for
square root and Procrustes there is clearly a large distance between {\it Lady Susan} and the rest, where {\it Lady
Susan} is her earliest work, a short novella published 54 years after Austen's death.
  
Figure \ref{ClusterMDS} (right column) shows corresponding plots of the first two multi-dimensional scaling (MDS)
variables from a classical multi-dimensional scaling analysis.  
The square root and Procrustes MDS plots are visually identical,
although they are slightly different numerically. We see that there is a clear separation in MDS space between
Austen's and Dickens' works with a very strong separation in MDS1 using the square root and Procrustes distances,
and less so for Euclidean distance. This example clearly shows differences when using the metrics, and demonstrates 
an advantage of using the square root Euclidean and Procrustes distances compared to the Euclidean distance here. 

%\textcolor{red}{We also compare the Euclidean metric used on the whole word set, this gives very similar results to using the $m=1000$ word set, the only difference is {\it David Copperfield} and {\it Great Expectations} are now clustered correctly %with the rest of Dickens work.}

\section{Framework for the statistical analysis of graph Laplacians}
\label{sec:space}
 \subsection{Framework}
\begin{figure}[htbp]
\centering
\begin{tikzpicture}[shorten >=1pt,node distance=3cm,on grid,auto]
    \node[rectangle] (q_0) {};
    \node[rectangle] (q_1) [right=of q_0] { };
    \node[rectangle] (q_2) [right=of q_1] {$Image(\mathcal{L}_m)$};
	\node[rectangle] (q_3) [right=of q_2] {};

    \node[rectangle] (q_4) [above =0.8cm  of q_0] {};
    \node[rectangle] (q_5) [above =0.8cm  of q_1]{ };
    \node[rectangle] (q_6) [right=of q_5] {$\mathcal{M}_m$};
	\node[rectangle] (q_7) [right=of q_6] {};
	\node[rectangle] (q_8) [above =0.4cm  of q_0] {$T_\nu(\mathcal{M}_m)$};
	\node[rectangle] (q_9) [above =0.4cm  of q_3] {$\mathcal{L}_m$}; 
		\node[rectangle] (q_10) [above =0.4cm  of q_1] { $\mathcal{M}_m$};
			\node[rectangle] (q_11) [above =0.4cm  of q_2] {\rotatebox[origin=c]{90}{$\subset$}};
	
    \path[->]
    (q_2) edge [bend left] node {$Id$} (q_1)
    (q_1) edge [bend left] node {$\pi_\nu^{-1}$} (q_0)
    (q_3) edge [bend left] node {$\text{F}_\alpha$} (q_2)
    %(q_1) edge [supset] node {} (q_0)
    (q_6) edge [bend left] node {$\text{P}_\mathcal{L}$} (q_7)
    %(q_2) edge [supset] node {} (q_1)
    (q_5) edge [bend left] node {$\text{G}_\alpha$} (q_6)
    (q_4) edge [bend left] node {$\pi_\nu$} (q_5);
    %(q_9) edge [supset] node {} (q_11);
\end{tikzpicture} \caption{\footnotesize{\textit{{\textit{Schematic diagram for the general framework for the statistical analysis of graph Laplacians. The embedding map $\text{F}_\alpha$ and embedding space $\mathcal{M}_{m}$ are defined in  Section \ref{sec:embed}.   The identity map is denoted by $Id$. The tangent space, $T_{\nu}(\mathcal{M}_m)$ and associated projections $\pi_{\nu}$ and $\pi_{\nu}^{-1}$ are defined in Section \ref{sec:tangent space}. 
The reverse power map $\text{G}_\alpha$ is  defined in Section \ref{reverse} and 
the projection $\text{P}_\mathcal{L}$ is defined in Section \ref{sec:project}.}}}}} \label{fig: projection}
\end{figure}
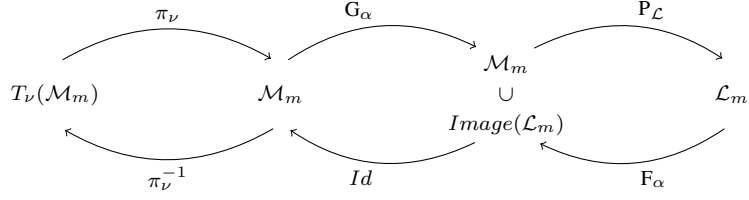

The general framework we will define in this section for the statistical analysis of graph Laplacians involves 
mapping, embedding and projections, shown schematically in Figure \ref{fig: projection}.

Distance metrics such as (\ref{eq:euc power metric}) and (\ref{eq:proc power metric}) on manifolds are referred
to as \emph{intrinsic} or \emph{extrinsic}.   An intrinsic distance is the length of a shortest geodesic path in the
manifold, whereas an extrinsic distance is one induced by a Euclidean distance in an embedding of the manifold
\citep[p112]{MR3559734}.  On $\mathcal{L}_m$, Euclidean distance $d_1$ is intrinsic, but in general $d_\alpha$ and
$d_{\alpha,S}$ are extrinsic with respect to an embedding defined as follows.
  
  \subsection{Map and embedding}\label{sec:embed}
We write $\textbf{L}=\textbf{U}\boldsymbol{\Xi}\textbf{U}^{T}$ by the
spectral decomposition theorem, with $\boldsymbol{\Xi}=\text{diag}(\xi_1,\ldots,\xi_m)$ and
$\textbf{U}=(\textbf{u}_1,\ldots,\textbf{u}_m)$, where $\lbrace\xi_i\rbrace_{i=1,\ldots,m}$ and
$\lbrace\textbf{u}_i\rbrace_{i=1,\ldots,m}$ are the eigenvalues and corresponding 
eigenvectors of $\textbf{L}$. We consider the following map which raises the graph Laplacian to the power $\alpha>0$:
\begin{align}
\begin{split}
\text{F}_\alpha(\textbf{L})&= \textbf{L}^\alpha=\textbf{U}\boldsymbol{\Xi}^\alpha\textbf{U}^T : \mathcal{L}_m\rightarrow Image(\mathcal{L}_m) \subset \mathcal{M}_m.\\
\end{split}  \label{Fmap}
\end{align}
We note that $\text{F}_\alpha$ is a bijective map (with inverse $\text{F}^{-1}_\alpha$). After applying the transformation  $\text{F}_\alpha$   we then consider the image of the graph Laplacian space to be embedded in a manifold ${\cal M}_m$, where statistical analysis is carried out
using extrinsic methods. We will either use the Euclidean distance $d_E$ (\ref{Euc}) or the Procrustes distance $d_S$ (\ref{Proc}) in the embedding manifold ${\cal M}_m$. 

Our choice of ${\cal M}_m$ will depend on which metric is used. 
When using the Euclidean power distance we take the embedding manifold ${\cal M}_m$ to be the space of real symmetric $m \times m$ matrices 
with centred rows and columns 
\begin{align}\label{eq: Mm define}
\mathcal{S}^*_m=\lbrace \textbf{Y}=(y_{ij}):\textbf{Y}=\textbf{Y}^T ;\, \textbf{Y} {\textbf{1}_m}={\textbf{0}_m} \rbrace,
\end{align}
which has dimension $m(m-1)/2$, which is the same dimension as $\mathcal{L}_m$. 
When using the Procrustes power distance we take the embedding manifold to be the reflection size-and-shape space \citep[p67]{10.2307/30242879,MR3559734}
\begin{align}\label{eq: Mm define2}
RS\Sigma^m_{m-1} = \{  \mathbb{R}^{(m-1)^2}  / O(m-1) \} ,
\end{align}  
which also has dimension $m(m-1)/2$. The reflection size-and-shape space has singularities, but away from these singularity sets the space is a Riemannian manifold.  

The distance metrics (\ref{eq:euc power metric}) and (\ref{eq:proc power metric}) are isometric to 
Euclidean distance in $\mathcal{S}^*_m$ and Procrustes distance in $RS \Sigma^m_{m-1}$ 
respectively, and can be written as
\begin{align*}
d_\alpha(\textbf{L}_1, \textbf{L}_2)&=\Vert\text{F}_\alpha(\textbf{L}_1)-\text{F}_\alpha(\textbf{L}_2)\Vert \\
d_{\alpha,S}(\textbf{L}_1, \textbf{L}_2)&=\inf_{\textbf{R}\in \mathcal{O}(m)}\Vert\text{F}_\alpha(\textbf{L}_1)-\text{F}_\alpha(\textbf{L}_2)\textbf{R}\Vert.
\end{align*}
%These are extrinsic distances, except for $d_1$.  
%These distances in fact hold more generally for $\textbf{L}_1, \textbf{L}_2 \in \mathcal{PSD}_m$ defined in (\ref{PSD}). 

A suitable choice of $\alpha>0$ will be dependent on the application. Some discussion in related work on symmetric positive semi-definite matrices is relevant. 
Despite the swelling effect that can be present for $\alpha=1$ an advantage in this case is that we have an 
intrinsic distance, and the nodes are directly used in the distance calculations rather than in an embedding space. 
The parameter $\alpha$ behaves like a Box-Cox transformation parameter, with $\alpha =\frac{1}{2}$ a matrix square root, 
and $\alpha \to 0$ a matrix logarithm. Large $\alpha$ gives strong weight to the differences between the largest values 
in the embedding space, and small $\alpha$ gives a more even weighting between large and small values. 
In \citet{Pigolietal14} the authors considered metrics between covariance operators including Procrustes metrics and in the continuous case it is required
that $\alpha \ge \frac{1}{2}$. In their examples $\alpha=\frac{1}{2}$ gave good results when investigating differences between
languages. \citet{Drydenetal10} also found  $\alpha=\frac{1}{2}$ was appropriate when using Box-Cox transformation for comparing 
diffusion weighted images. The Procrustes distance with $\alpha=\frac{1}{2}$ between two covariance operators
is the same as the Wasserstein distance between two zero mean 
Gaussian processes with different covariance operators \citep{Masarottoetal19}. The popularity of the Wasserstein metric 
as an optimal transport distance between probability distributions \citep[e.g.][Chapter 6]{Villani09} lends further support to using both Procrustes version and $\alpha=\frac{1}{2}$. 
However, ultimately it is of course up to the user whether to use Procrustes or not, and which $\alpha$ to choose. In our applications we will compare  
$\alpha=1$ and $\alpha = \frac{1}{2}$.

\subsection{Tangent space}\label{sec:tangent space}
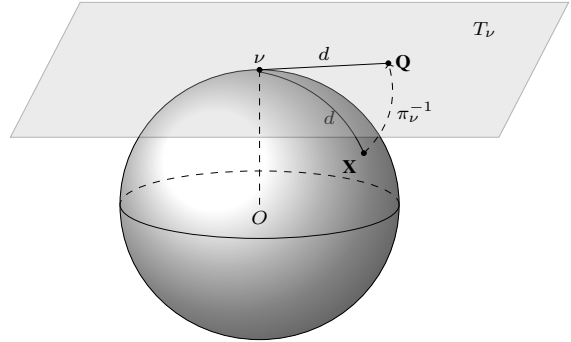
\begin{figure}[htbp]
\centering
\resizebox{7.5cm}{4.5cm}{%
\begin{tikzpicture}
[
  point/.style = {draw, circle, fill=black, inner sep=0.7pt},
]
\def\rad{2cm}
\coordinate (O) at (0,0); 
\coordinate (N) at (0,\rad);
\coordinate (R) at (0.923*\rad-10,0.382*\rad); %X
\coordinate (TR) at (0.923*\rad,1.4*\rad-20);%Q

\filldraw[ball color=white] (O) circle [radius=\rad];
\draw[dashed] 
  (\rad,0) arc [start angle=0,end angle=180,x radius=\rad,y radius=5mm];
\draw
  (\rad,0) arc [start angle=0,end angle=-180,x radius=\rad,y radius=5mm];
  \draw[black]
  (R) arc [start angle=22.5,end angle=80,x radius=\rad,y radius=20mm];
  \node[black] at (1,1.3) {$d$};  
\begin{scope}[xslant=0.5,yshift=\rad,xshift=-2]
\filldraw[fill=gray!50,opacity=0.3]
  (-4,1) -- (3,1) -- (3,-1) -- (-4,-1) -- cycle;
\node at (2,0.6) {$T_{{\nu}}$};  
\end{scope}
\draw[dashed]
  (N) node[above] {$\nu$} -- (O) node[below] {$O$};
\draw[black]
  (N) node[above] {} -- node[above] {$d$}(TR) node[right] {};
\draw (R)node[below left]{$\textbf{X}$};
\draw (TR)node[right]{$\textbf{Q}$};
\path[->]
(R) edge [bend right] node [right]{$\pi_\nu^{-1}$} (TR)[dashed];
\node[point] at (N) {};
\node[point] at (R) {};
\node[point] at (TR) {};
\end{tikzpicture}
}
\caption{\footnotesize{\textit{A diagrammatic view of a $\pi_\nu^{-1}$ projection, mapping $\textbf{X}$ 
from a manifold ${\mathcal M}_m$, here shown schematically as a sphere, 
onto the tangent space $T_\nu$.}}}\label{fig:tangent space}
\end{figure}

To perform statistical analysis we work with a tangent space at pole $\nu \in \mathcal{M}_{m}$ which we denote by $T_\nu(\mathcal{M}_{m})$. 
A projection from the tangent space $T_\nu(\mathcal{M}_{m})$ to $\mathcal{M}_{m}$ is written as
$$ \pi_\nu :  T_\nu(\mathcal{M}_{m}) \to \mathcal{M}_{m} $$ 
with inverse projection $\pi_\nu^{-1}$. Standard statistical methods can be applied in the tangent space, which is a Euclidean space of dimension $m(m-1)/2$. 
Figure \ref{fig:tangent space} shows a simple visualisation of a tangent space. 
The tangent space at $\nu$ is a Euclidean approximation touching the manifold $\mathcal{M}_{m}$. In non-Euclidean spaces a distance is the length of the shortest geodesic path between two points on a manifold. For specific Riemannian 
metrics the tangent projection could be the inverse exponential map, denoted $\exp^{-1}_{\nu}$ \citep[Chapter 5]{MR3559734}, 
and in this case a geodesic becomes a straight line in the tangent space preserving distance to the pole.

As the graph Laplacian space has centering constraints on the rows and columns, these constraints are also 
preserved in our choice of map and embedding to ${\cal M}_m$. We can remove the centering constraints and reduce the dimensions  
of the matrices when projecting to a tangent space by pre and post 
multiplying by the $m-1 \times m$ Helmert sub-matrix $\textbf{H}$ and its transpose as a component of the projection. The 
Helmert sub matrix $\textbf{H}$, of dimension $m-1\times m$, has $j$th row defined as
\begin{align*}
(\underbrace{h_j,\ldots,h_j}_{j \text{ times}},-jh_j,\underbrace{0,\ldots,0}_{m-j-1 \text{ times}}), \quad h_j=-(j(j+1))^{-\frac{1}{2}},
\end{align*}
(page 49, \citet{MR3559734}).  Note that $\textbf{H}\textbf{H}^T = \mathbf{I}_{m-1}$ and 
$\textbf{H}^T\textbf{H} = \mathbf{C}_m$,  where $\mathbf{C}_m = \mathbf{I}_m - \mathbf{1}_m\mathbf{1}_m^T/m$ is the $m \times m$ centering matrix, $\mathbf{I}_m$ is
the $m \times m$ identity matrix and $\mathbf{1}_m$ is the $m$-vector of ones.

For the Euclidean power metric we use the inverse tangent projection  $\pi_\nu^{-1}$   to 
the tangent space $T_\nu({\cal S}_m^*) = \mathbb{R}^\frac{m(m-1)}{2} $ as 
\begin{align}
\begin{split}
\pi_\nu^{-1}(\textbf{X})=\text{vech*}\{ \textbf{H} (\textbf{X}-\nu)  \textbf{H}^T \}\\
\pi_\nu(\textbf{Q})=   \nu + \textbf{H}^T  \text{vech*}^{-1} ( \textbf{Q} )    \textbf{H} 
\end{split}
\end{align}
where ${\rm vech}^*$ is the half vectorisation of a matrix including the diagonal, similar to ${\rm vech}$,  but with $\surd 2$ multiplying the elements  corresponding to the off-diagonal.
In this case $\mathcal{M}_m = \mathcal{S}_m^*$ as in (\ref{eq: Mm define}) so has zero curvature, and
analysis is unaffected by the choice of $\nu$. The use of the Helmert sub-matrix ensures that we 
have the correct number $m(m-1)/2$ of tangent coordinates, and is a convenient way of dealing with the centering constraints in   $\mathcal{M}_m$.

For the Procrustes power metric we define the map $\pi_\nu^{-1}$ to 
the tangent space\\ $T_\nu(RS \Sigma^m_{m-1}) = \mathbb{R}^\frac{m(m-1)}{2}$ as
\begin{align}
\begin{split}
\pi_\nu^{-1}(\textbf{X})= \text{vec} \{ \textbf{H}(\textbf{X}\hat{\textbf{R}}-\nu)\textbf{H}^T  \}\\
\pi_\nu(\textbf{Q})=  ( \nu + \textbf{H}^T \text{vec}^{-1} (\textbf{Q}) \textbf{H}   ) \tilde{\textbf{R}}  
\end{split} \label{tanproc}
\end{align}
where $\rm{vec}$ is the vectorise operator obtained from stacking the columns of a matrix,  
{$\hat{\textbf{R}}$ is the ordinary Procrustes match of $\textbf{X}$ to $\nu$ \citep[chapter 7]{MR3559734} and 
$\tilde{\textbf{R}}$ is the ordinary Procrustes match from $( \nu+    \textbf{H}^T \text{vec}^{-1}(\textbf{Q}) \textbf{H}   )$ to $ \nu$. }  Note that the reflection size-and-shape 
space is a space with positive curvature \citep{Kendetal99} and statistical analysis depends on the choice of
$\nu$. A sensible choice for $\nu$ is the sample mean, as defined in Section \ref{samplemean}.

\subsection{Reverse power map}\label{reverse}
When transforming back from the embedding space to the graph Laplacian space we choose a practical method which involves 
first applying a continuous reverse power map $\textbf{G}_\alpha$ and then a projection $\text{P}_\mathcal{L}$ into graph Laplacian space:
$$\text{P}_\mathcal{L} \circ  \textbf{G}_\alpha  : \mathcal{M}_m \to \mathcal{L}_m , $$
as illustrated in the framework in Figure \ref{fig: projection}.

We consider four choices for the reverse power map:
\begin{align*}
\text{G}_\alpha(\textbf{Q})&=\begin{cases}
\quad (\textbf{Q})^\frac{1}{\alpha} \;  \; \text{ when } \frac{1}{\alpha} \text{ is an odd integer}&  : \mathcal{M}_{m} \rightarrow\mathcal{M}_m  \\
\quad \left( \frac{\textbf{Q}+\textbf{Q}^T+ \{ (\textbf{Q}+\textbf{Q}^T)^T(\textbf{Q}+\textbf{Q}^T) \} ^\frac{1}{2}}{4}\right)^\frac{1}{\alpha} & :   \mathcal{M}_{m} \rightarrow\mathcal{PSD}_{m}^* \subseteq \mathcal{M}_m\\
\quad (\textbf{QQ}^T)^\frac{1}{2\alpha}  & :  \mathcal{M}_m \rightarrow\mathcal{PSD}_{m}^*  \subseteq \mathcal{M}_m,\\
\quad (\textbf{Q}^T\textbf{Q})^\frac{1}{2\alpha}  & :  \mathcal{M}_m \rightarrow\mathcal{PSD}_{m}^*  \subseteq \mathcal{M}_m.
\end{cases}
\end{align*}
which are suitable for different scenarios
depending on whether or not we want to map to the space of centred positive semi-definite matrices 
$\mathcal{PSD}_m^*$ before reversing the powering of $\alpha$. 
In our applications we use the first choice of reverse map for Euclidean distance $d_1$, which is just the identity map in this case. The second expression before taking the power $\frac{1}{\alpha}$ is the closest symmetric positive semi-definite matrix to $\textbf{Q}$ in terms of Frobenius distance \citep{higham1988computing}, and we use this reverse map for $d_{\frac{1}{2}}$.  The third reverse map removes the orthogonal matrix from the Procrustes match introduced from the tangent projection in (\ref{tanproc}) and was used previously by \citet{10.2307/30242879}. We use the fourth reverse map for  $d_{\frac{1}{2},S}$ in our applications, where the orthogonal matrix from the Procrustes match is retained.

\subsection{Projection}\label{sec:project}
Suitable choices of projection $\text{P}_\mathcal{L}$ based on the Euclidean power or Procrustes power metrics are:  
\begin{align}
\begin{split}
\text{P}_\alpha(\textbf{Y})&=  \arg\inf_{\textbf{L}\in\mathcal{L}_m}  d_E( \textbf{Y} , F_\alpha(\textbf{L}) )   : \mathcal{M}_{m}\rightarrow\mathcal{L}_{m}  \\
\text{P}_{\alpha,S}(\textbf{Y})&=  \arg\inf_{\textbf{L}\in\mathcal{L}_m}   d_S( \textbf{Y} , F_\alpha(\textbf{L})   )  : \mathcal{M}_{m}\rightarrow\mathcal{L}_{m} ,
\end{split}
\end{align}
where the Euclidean distance $d_E$ and the Procrustes distance $d_S$ were defined in (\ref{Euc}) and (\ref{Proc}), respectively.  

It is desirable that optimisation involved in computing the projection is convex, 
since convex optimisation problems have the
useful characteristic that any local minimum must be the unique global minimum 
\citep{rockafellar1993lagrange}.
  
\begin{result}\label{theorem: convex proj}
For $\text{P}_1$ the projection can be found by solving a convex optimisation problem with a unique solution, by minimising
\begin{align}\label{eq: euc optimisation}
\begin{split}
f(\textbf{Y})&=d_E^2(\textbf{L},\textbf{Y})=\sum_{i=1}^m\sum_{j=1}^m(l_{ij}-y_{ij})^2\\
 \text{subject to:}\quad \quad l_{ij}-l_{ji}&=0, \quad 0\leq i, j\leq m\\
\sum_{j=1}^ml_{ij}&=0, \quad 0\leq i\leq m\\
l_{ij} &\leq 0, \quad \quad 0\leq i, j\leq m \text{ and } i\neq j.
\end{split}
\end{align}
\end{result}
It is immediately clear that this is a convex optimization problem since the objective function is quadratic with 
Hessian $2\mathbf{I}_{m(m-1)/2}$, which is strictly positive 
definite, and the constraints are linear.  
The unique global solution for the projection can be found using quadratic programming.
To implement this projection $\text{P}_1$ we can, for example,  use either the {\tt CVXR} \citep{CVXR18} or {\tt
rosqp} \citep{rosqp18} packages in R \citep{CRAN18} to solve the optimisation, and {\tt rosqp} is 
particularly fast even for $m =1000$. 

Note that the choice of metric for projection does not need to be the same as the choice of metric for estimation. 
As the projection for the Euclidean power metric with $\alpha=1$ involves convex optimisation we will use $\text{P}_\mathcal{L} = \text{P}_1$ throughout for all our metrics. For $\alpha \ne 1$ the optimization is not in general convex.

An alternative procedure to using the reverse map followed by a projection could be to first apply a projection into the image space of graph Laplacians 
and then apply an inverse of the embedding map $\textbf{F}^{-1}_\alpha$, in a similar manner to \citet{lin2017extrinsic}. 
This alternative approach is more difficult to work with in general for graph Laplacians, except for the Euclidean metric with $\alpha=1$ when
both approaches are equivalent.

\subsection{Mean estimation}\label{samplemean}
There are two main types of means on a manifold: an intrinsic mean and an extrinsic mean \citep[Chapter 6]{MR3559734}. 
 We define the mean in the graph Laplacian space using extrinsic means, although the mean when the Euclidean power distance with $\alpha=1$ is used is in fact an intrinsic mean. 

 We define the population mean for graph Laplacians as
\begin{align}\label{eq: pop mean}
\begin{split}
\mu= \text{P}_\mathcal{L} (\textbf{G}_{\alpha}(\eta)),\text{ where }{\eta}= {\rm arg} \inf_{u\in \mathcal{M}_m}\textbf{E}[d^2(\textbf{F}_{\alpha}(\textbf{L}), u)],
\end{split}
\end{align}
assuming $\eta$ exists, and $d$ is either the Euclidean or Procrustes distance in $\mathcal{M}_m$. 
We define the sample mean for a set of graph Laplacians as
\begin{align}\label{eq:sample mean}
\begin{split}
\hat{\mu}=\text{P}_\mathcal{L} (\textbf{G}_{\alpha}(\hat{\eta})),\text{ where }
\hat{\eta}= {\rm arg} \inf_{u\in \mathcal{M}_m}\frac{1}{n}\sum_{k=1}^n d^2(\textbf{F}_{\alpha}(\textbf{L}_k), u),
\end{split}
\end{align}
assuming $\hat\eta$ exists. 
For the Euclidean power distance we have 
\begin{align}
{\eta}&= \mathbb{E}[\text{F}_\alpha(\textbf{L})]  \label{eta1}\\
\hat{\eta}&= \frac{1}{n}\sum_{k=1}^n\text{F}_\alpha(\textbf{L}_k),  \label{eta2}
\end{align}
where  $\eta, \hat\eta$, and hence $\mu, \hat{\mu}$, are unique in this case.   For the Euclidean power metric when $\alpha=1$,  we have $\hat{\mu}=\hat{\eta}$ and the mean is a Fr{\'e}chet intrinsic mean \citep{Frechet1948, ginestet2017hypothesis} in this case.  An alternative discussion of the Fr\'echet mean is given by \citet{kolaczyk2020} who use vectorization rather than the matrices that we use. 

For the Procrustes power distance $\mu$ and $\hat{\mu}$ may be sets,  and the conditions for uniqueness rely on the curvature of the space 
 \citep{10.2307/1428094}. 
 %In particular the support of the distribution is a geodesic ball $B_r$ such that $B_{2r}$ is regular.  
 We will assume uniqueness exists throughout. In the Procrustes case we assume uniqueness up to orthogonal transformation.

 \begin{result}\label{theorem:euc mean}
 Let $\textbf{L}_k, \; k=1,\ldots,n$ be a random sample of i.i.d. observations from a distribution with population mean $\mu$ in (\ref{eq: pop mean}). 
 For the power Euclidean distance ${d}_\alpha$ the estimator $\hat{\mu}$, in (\ref{eq:sample mean}), is a consistent estimator of $\mu$. 
 \end{result}
 The proof of this result can be found in Appendix \ref{sec:cals for space}. Note that a similar result holds for ${d}_{\alpha,S}$ where stronger conditions for consistency of $\hat{\eta}$ are given in \citet{bhattacharya2003}, 
 but the same projection argument used in the proof  holds.

\begin{figure}[htbp]
\includegraphics[width=12.5cm]{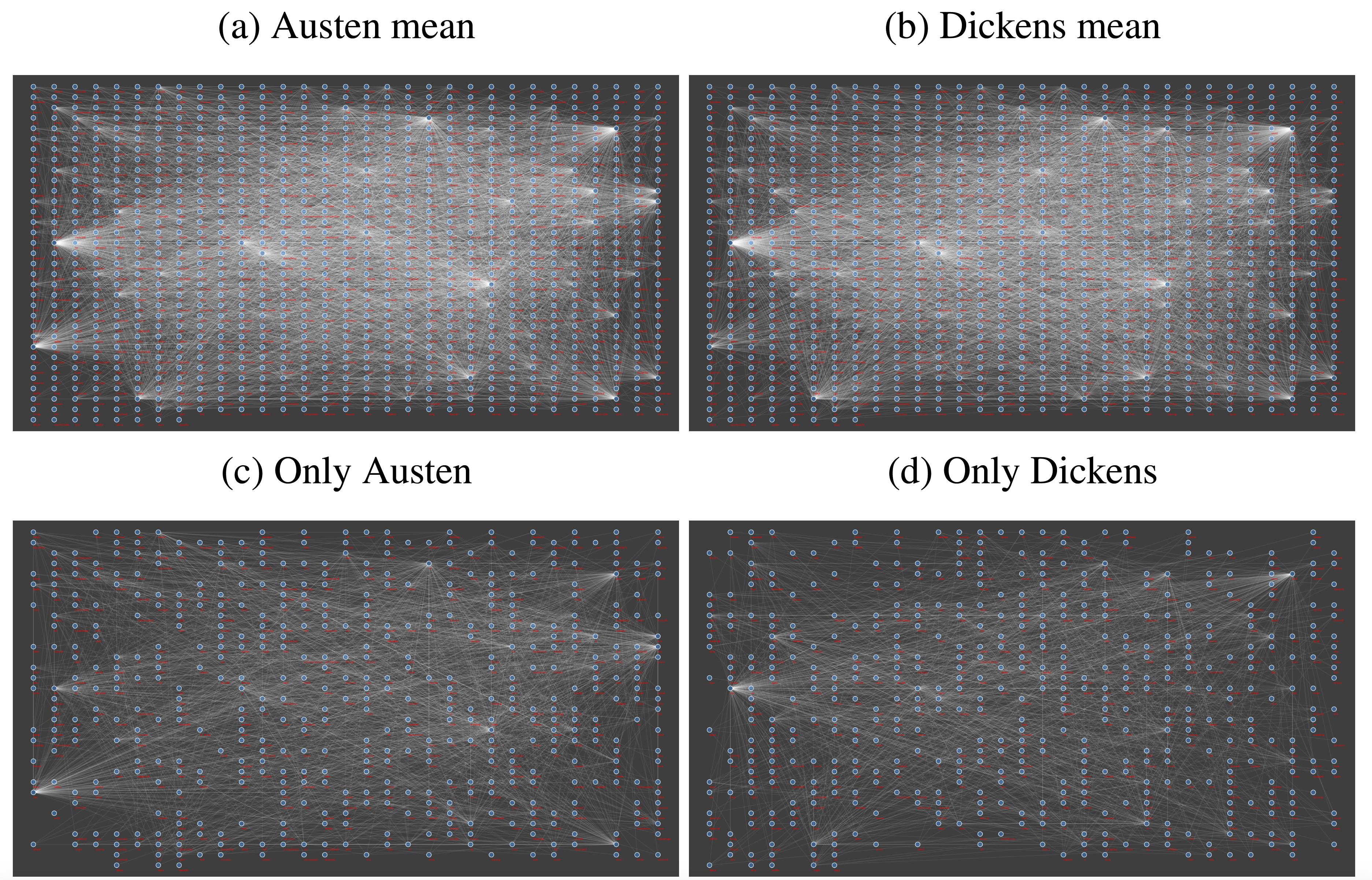}
\caption{ \footnotesize{\textit{The means of (a) Austen's novels and (b) Dickens' novels using $d_1$ based on the top m=1000 word pairs. In (c) we see edges present in the
Austen mean but not Dickens and in (d) the edges present in Dickens and not Austen means. Zoom in for more detail. }}}\label{hugeplots}        
\end{figure}

Figure \ref{hugeplots} shows an illustration of the sample means for Austen and Dickens novels using $d_1$, with the 1000 words arranged in a grid and edges drawn between words which co-occur with adjacency weight at least $10^{-5}$ of the sum of the nodes. 
The means plots for both authors are similar, perhaps unsurprisingly as approximately half of the words 
in each novel are represented by the first 50 words.  
Figure (c) shows edges present in the Austen mean but not in the Dickens mean, and (d) 
the edges present in the Dickens mean but not in the Austen mean, to highlight the differences between the two networks.  
By zooming in, the mean plots illustrate more co-occurrences involving {\it she}, {\it her} by Austen and involving {\it the}, {\it his}, {\it don't}  by Dickens, among many others. These plots are drawn using the program {\tt Cytoscape} 
\citep{Cytoscape03}. We shall explore the differences in more detail later in Section \ref{explore}.  Alternative plots comparing the means using the Euclidean, the square root Euclidean and Procrustes metric can be found in Appendix \ref{sec:other means}, and all are visually very similar.

\subsection{Interpolation and extrapolation}
We now consider an interpolation path, \\$\textbf{L}(c)$, for $c$ being the position along the path,  $0\leq c \leq 1$,  between the graph Laplacians at $\textbf{L}(0)$ and $\textbf{L}(1)$. For $c<0$ and $c>1$ the path $\textbf{L}(c)$ is extrapolating from the graph Laplacians, at $\textbf{L}(0)$ and $\textbf{L}(1)$. 
The interpolation and extrapolation path between graph Laplacians for each metric is defined by first finding the geodesic path in the embedding space ${\cal M}_m$ between the embedded graph Laplacians, which is then projected to $\mathcal{L}_m$. 
 
The interpolation and extrapolation path passing through $\textbf{L}_1=\text{P}_\mathcal{L}(\text{G}_\alpha(\nu))$ and $\textbf{L}_2$ is
\begin{align}
\textbf{L}(c)=\text{P}_\mathcal{L}(\text{G}_\alpha(\pi_\nu \lbrace c \pi_\nu^{-1} (\text{F}_\alpha(\textbf{L}_2))\rbrace)).
\end{align} 
 
For the Euclidean power this simplifies to
 \begin{align}
 \textbf{L}(c)=\text{P}_\mathcal{L}(\text{G}_\alpha(\text{F}_\alpha(\textbf{L}_1)+c(\text{F}_\alpha(\textbf{L}_2)-\text{F}_\alpha(\textbf{L}_1)))).
 \end{align}
 
Although these paths are geodesics in ${\cal M}_m$ they may not be geodesics when mapped back to the
graph Laplacian space.  For the Euclidean power case with $\alpha=1$ the distance $d_1$ is an intrinsic distance and the interpolation paths are minimal geodesics 
given by \begin{align}
\textbf{L}(c)= (1-c)\textbf{L}_1+c\textbf{L}_2\in \mathcal{L}_m \; \; \; \; 0 \le c \le 1.
\end{align}
But for $\alpha \ne 1$ and the Procrustes power metrics for any $\alpha$ the distances are extrinsic.

\begin{figure}[htbp]

    \centering
   \begin{subfigure}[b]{0.3\textwidth}
\caption{$c=-5$ }
        \includegraphics[width=3.3cm,trim={2cm 0 2cm 2cm}, clip,scale=0.45]{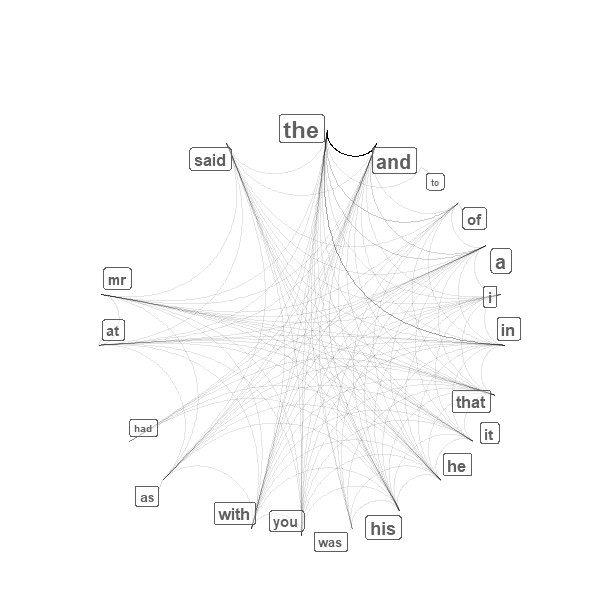}
    \end{subfigure}
             \begin{subfigure}[b]{0.3\textwidth}
\caption{ $c=0.5$}
        \includegraphics[width=3.3cm,trim={2cm 0 2cm 2cm}, clip,scale=0.45]{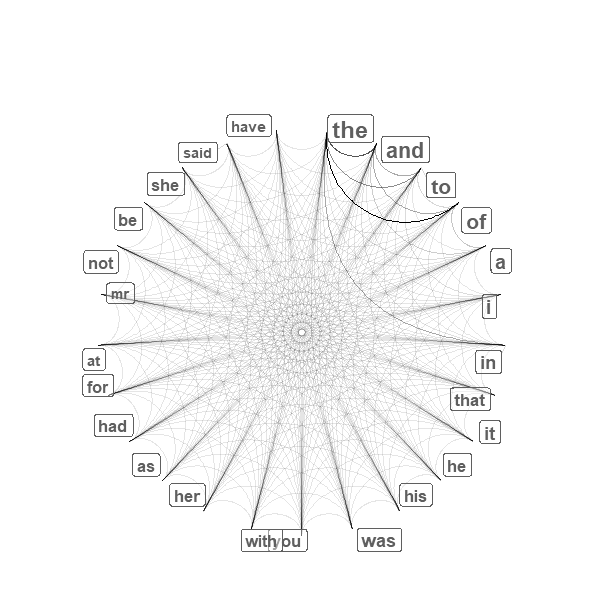}
    \end{subfigure}   
          \begin{subfigure}[b]{0.3\textwidth}
\caption{ $c=6$}
        \includegraphics[width=3.3cm,trim={2cm 0 2cm 2cm}, clip,scale=0.45]{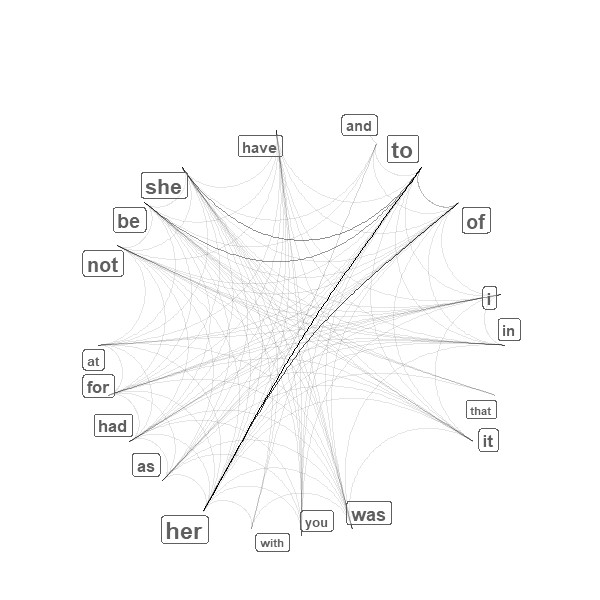}
    \end{subfigure}
              \begin{subfigure}[b]{0.3\textwidth}
\caption{ $c=-5$}
        \includegraphics[width=3.3cm,trim={2cm 0 2cm 2cm}, clip,scale=0.45]{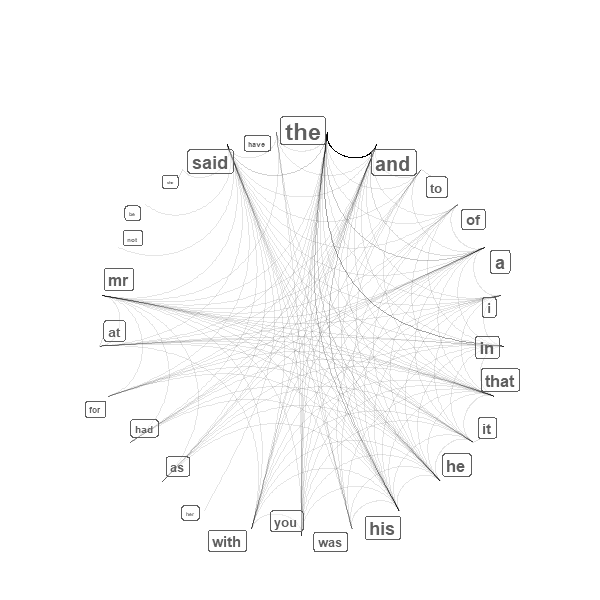}
    \end{subfigure}
    \begin{subfigure}[b]{0.3\textwidth}
\caption{ $c=0.5$}
     \includegraphics[width=3.3cm,trim={2cm 0 2cm 2cm}, clip,scale=0.45]{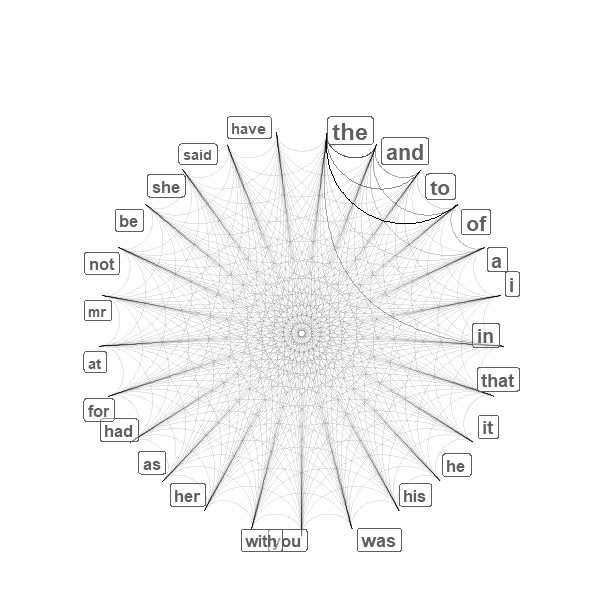}
    \end{subfigure}   
          \begin{subfigure}[b]{0.3\textwidth}
\caption{ $c=6$}
        \includegraphics[width=3.3cm,trim={2cm 0 2cm 2cm}, clip,scale=0.45]{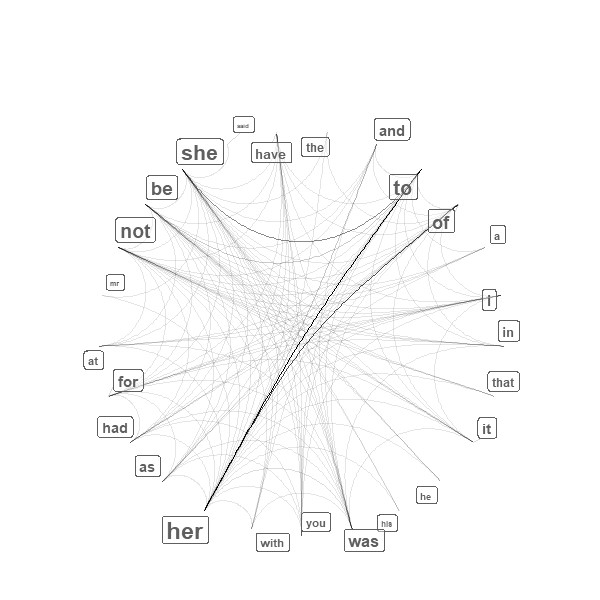}
    \end{subfigure}
                  \begin{subfigure}[b]{0.3\textwidth}
\caption{ $c=-5$}
        \includegraphics[width=3.3cm,trim={2cm 0 2cm 2cm}, clip,scale=0.45]{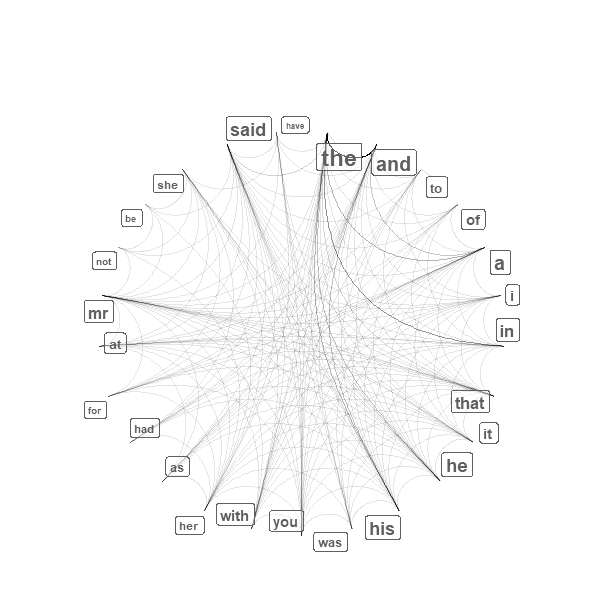}
    \end{subfigure}
    \begin{subfigure}[b]{0.3\textwidth}
\caption{ $c=0.5$}
     \includegraphics[width=3.3cm,trim={2cm 0 2cm 2cm}, clip,scale=0.45]{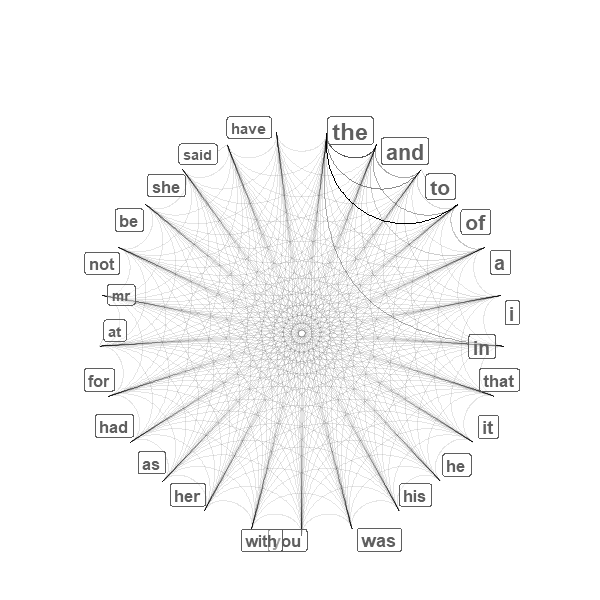}
    \end{subfigure}   
          \begin{subfigure}[b]{0.3\textwidth}
\caption{ $c=6$}
        \includegraphics[width=3.3cm,trim={2cm 0 2cm 2cm}, clip,scale=0.45]{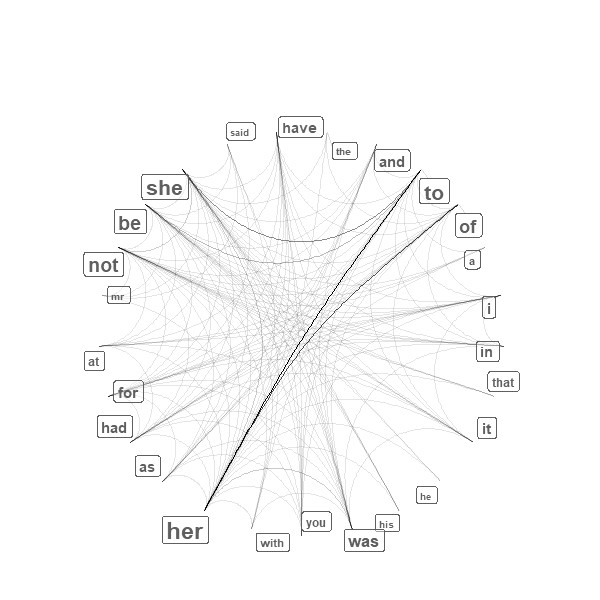}
    \end{subfigure}
     \caption{ \footnotesize{\textit{Networks on the interpolation and extrapolation paths between Dickens' and Austen's mean novels at  $c=0.5$ (interpolation) and $c=-5, c=6$ (extrapolations).  Different metrics are used in each row (top to bottom) $d_1$, $d_\frac{1}{2}$ and $d_{\frac{1}{2},S}$.     The top 25 words are displayed where the mean novels for the authors are estimated
     for each metric respectively using $m=1000$.}}}\label{fig:int extr}
        \end{figure}

Figure \ref{fig:int extr} shows networks evaluated on the interpolation and extrapolation paths at $c \in \{ -5,0.5,6 \}$ between the mean Austen and Dickens novels when using different metrics for the 25 nodes corresponding to the most frequent words out of $m=1000$ nodes. At $c=6$ the feminine words have larger degrees and their edges have larger weights, for example {\it her} to {\it to}, {\it of}  and {\it she} to {\it to}. For $c=-5$ the nodes for {\it she} and {\it her} are actually removed indicating they have degree 0, {which is further evidence of the fact Austen used female words more then Dickens.} The different choices of metrics lead to similar interpolation and extrapolation paths in this example, although the $d_1$ extrapolations are more sparse than $d_{\frac{1}{2}}$, which in turn are
more sparse than $d_{\frac{1}{2},S}$.

\section{Further inference}\label{sec:pca regression}

 \subsection{Principal component analysis}

    There are several generalisations of PCA to manifold data, and the approach we define is similar to \citet{fletcher2004principal}, by computing PCA in a tangent space and projecting back to the manifold. 
     Our approach differs from \citet{fletcher2004principal} in that we have the extra layer of embedding the manifold and in addition we apply the reverse map and 
 projection back to graph Laplacian space.  Earlier approaches 
    of PCA in tangent spaces in shape analysis include \citet{kent1994complex} and \citet{Cootetal94}.
   
      Let $\textbf{v}_k=(\pi_\nu^{-1} (\text{F}_\alpha(\textbf{L}_k)))$, where $\nu= \hat{\eta} $ for either the Euclidean or Procrustes power metric, then $\textbf{S}=\frac{1}{n}\sum_{k=1}^n\textbf{v}_k\textbf{v}_k^T$ is an estimated covariance matrix. 
Suppose \textbf{S} is of rank $r$ with non-zero eigenvalues $\lambda_1,\ldots, \lambda_{r}$, then the corresponding eigenvectors $\boldsymbol{\gamma}_1,\ldots, \boldsymbol{\gamma}_{r}$ are the 
principal components (PCs) in the tangent space, and the PC scores are
\begin{align}
s_{kj}=\boldsymbol{\gamma}_j^T\textbf{v}_k, \quad \text{ for } k=1,\ldots,n, \quad j=1,\ldots,r.
\end{align}
The path of the $j$th PC in $\mathcal{L}_m$ is
\begin{align}
\textbf{L}(c)=\text{P}_\mathcal{L}(\text{G}_\alpha( \, \pi_\nu (c\lambda_j^\frac{1}{2}\boldsymbol{\gamma}_j)\,)), \quad c\in\mathbb{R}.
\end{align}

For the Euclidean case when $\alpha=1$ is chosen, the importance of the $i$th word in the principal component $\boldsymbol{\gamma}$ is given by 
\begin{align} \label{eq:importance of pc}
\frac{\pi_\nu (\boldsymbol{\gamma})_{ii}}{(\sum_{j=1}^{m} \pi_\nu (\boldsymbol{\gamma})_{jj})}, \text{ for } 1\leq i \leq m.
\end{align}

\begin{figure}[htbp]
    \begin{center}
         \includegraphics[width=4.1cm]{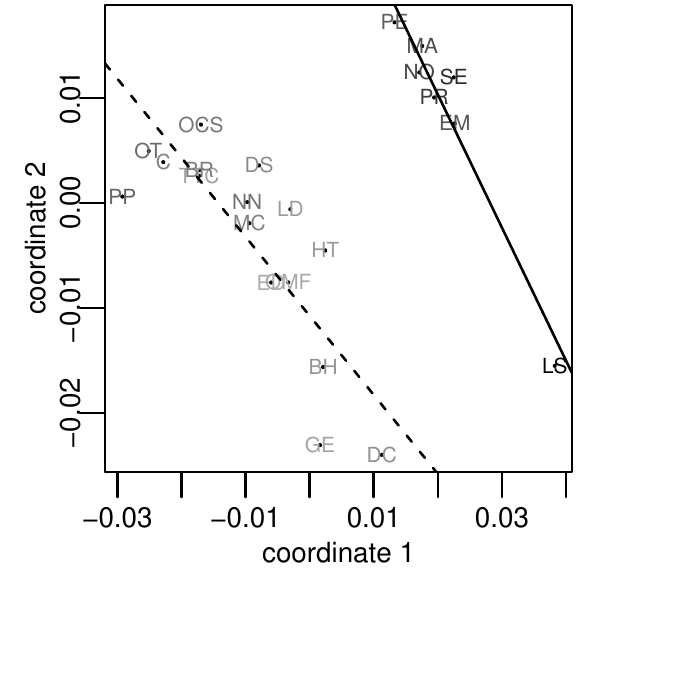}
         \includegraphics[width=4.1cm]{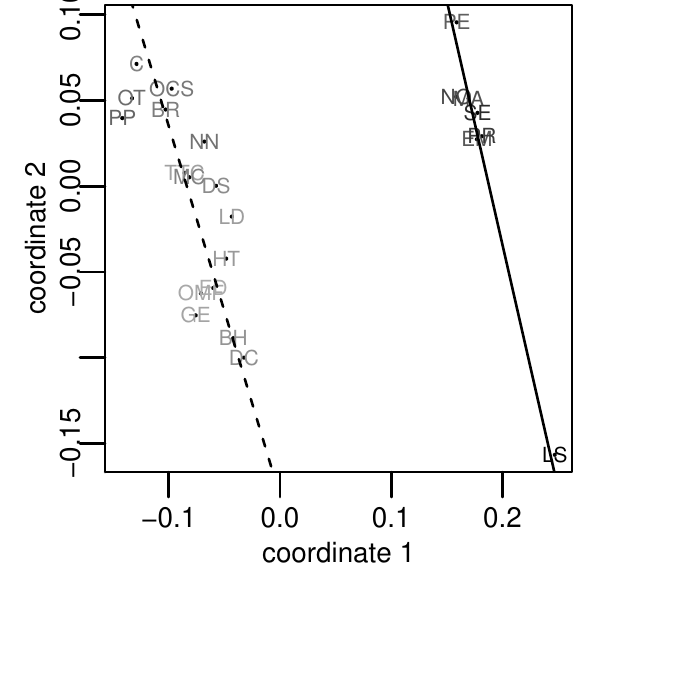}
         \includegraphics[width=4.1cm]{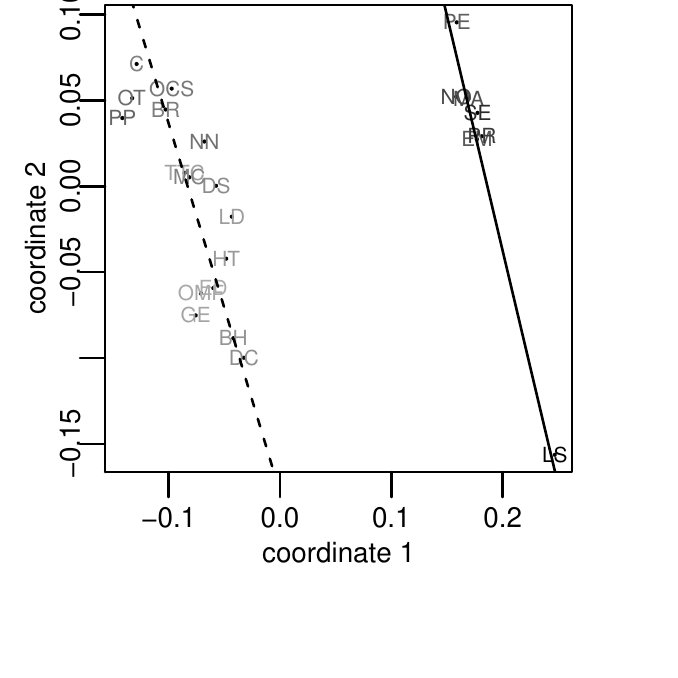}
   \end{center}
     \caption{ \footnotesize{\textit{Plot of PC 1 and PC 2 scores for the Austen and Dickens novels, shaded in time order (black to gray, exact times can be found in Table \ref{table:novels}) with extrinsic regression lines for Dickens novels (black) and Austen novels (dashed) using the (left) 
     Euclidean, (middle) square root Euclidean, and (right) Procrustes size-and-shape metric.}}}\label{fig:mds and PC COORDINATES}
        \end{figure}

We now apply the methods of PCA to the pooled samples of Austen and Dickens novels, for $m=1000$. The first and second PC scores are plotted in Figure \ref{fig:mds and PC COORDINATES} for the Euclidean, square root Euclidean and Procrustes metrics. Using the Procrustes metric gave visually identical results to the square root Euclidean as we have specified the 
labelling of the nodes using the most common words.  The extrinsic regression lines are included which we will define and explain below.  The variance explained by PC 1 and PC 1 and 2 together was 49$\%$ and 70$\%$; 37$\%$ and 46$\%$; and 37$\%$ and 46$\%$ for the Euclidean, square root Euclidean and Procrustes size-and-shape respectively. A benefit of the square root Euclidean and Procrustes metric is clear here as they separate the Austen and Dickens novels with a large gap on PC1 where as {\it David Copperfield} (DC) and {\it Persuasion} (PE) are very close in PC1 for the 
Euclidean case. We now analyse the Euclidean PCs in more detail.

\vskip 1cm 

 \begin{figure}[htbp]
    \centering
        \includegraphics[width=6cm,trim={0cm 0cm 0 0},clip]{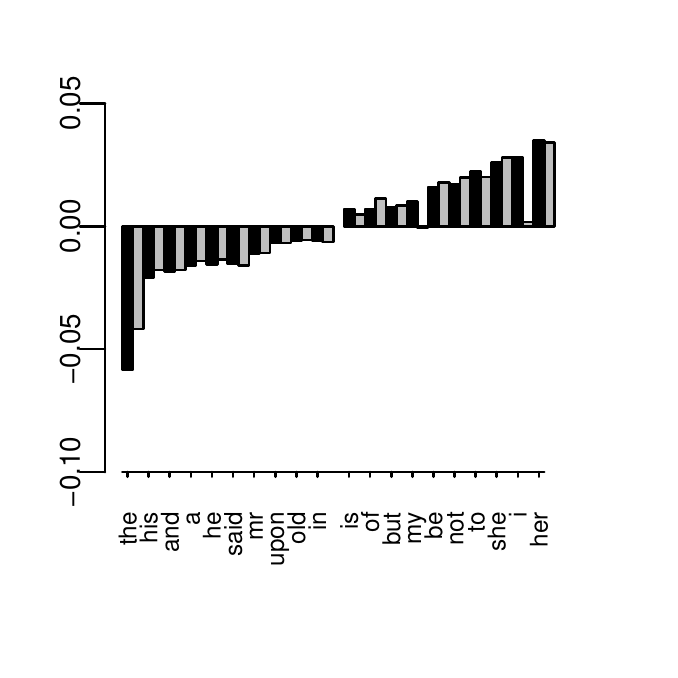}
        \includegraphics[width=6cm, trim={0cm 0cm 0 0},clip]{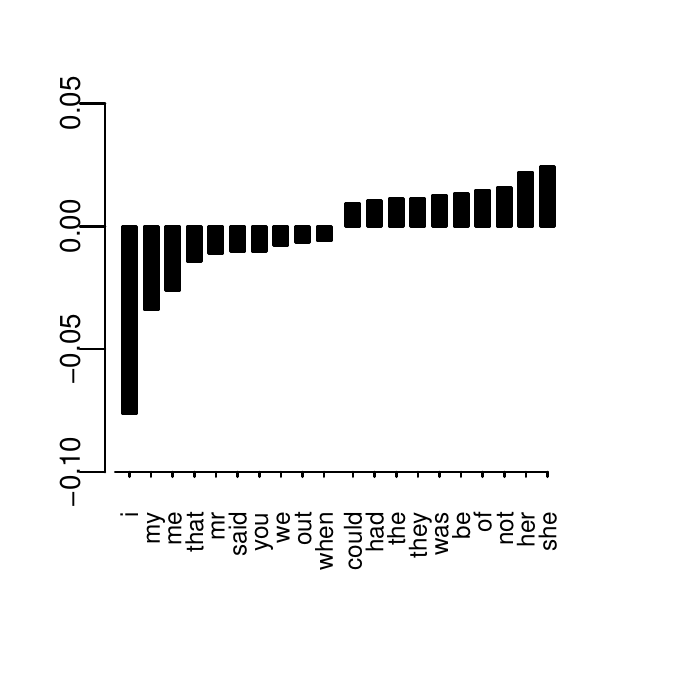}
            \caption{\footnotesize{\textit{The importance of each word given by (\ref{eq:importance of pc})
in (left) PC 1 and (right) PC 2. The gray bars on the left represent the importance of each word in the difference of means, $\textbf{D}=\hat{\mu}^{Austen}_E-\hat{\mu}^{Dickens}_E$, given by $((\textbf{D})_{ii})/({\sum_{j=1}^{1000} ({\textbf{D}})_{jj}})$, $1\leq i \leq 1000$.}}}\label{fig:euc PC degree}
        \end{figure}

        Figure \ref{fig:euc PC degree} contains plots representing the importance and sign of each word in the first and second Euclidean PC. 
          From Figure \ref{fig:mds and PC COORDINATES} a more positive PC 1 score is indicative of an Austen novel whilst a more negative one a Dickens novel. For a positive PC1 score the nodes {\it her} and {\it she} have importance whilst for a negative score words such as {\it his}, and {\it he} have more importance, which is expected as Austen writes with more female characters. 
          The second PC is actually similar to a fitted regression line which we describe in the next section and it is interesting to refer to the dates 
          that the novels were written in Table \ref{table:novels}. For Austen, the PC2 scores tend to be larger for later novels, and note that {\it Lady Susan} (LS) and  {\it Persuasion} (PE) are her earliest and latest novels respectively.  For Dickens the opposite is true, in that the PC2 scores tend to be smaller for later novels. {\it Pickwick papers} (PP) is Dickens' earliest and {\it The Mystery of Edwin Drood} (ED) his latest. 
The second PC has feminine words like {\it her} and {\it she} as the most positive words, but more first and second person words, such as {\it I}, {\it my} and {\it you} as negative words. This is consistent with Austen increasingly using a stylistic device called free indirect speech in her later novels \citep{10.2307/450561}. Free indirect speech has the property the third person pronouns, such as {\it she} and {\it her} 
are used instead of first person pronouns, such as {\it I} and {\it my}. 
%https://www.theguardian.com/books/2015/dec/05/jane-austen-emma-changed-face-fiction

 \subsection{Regression}\label{sec:regression} 
Here we assume the data are the pairs $\lbrace\textbf{L}_k, \textbf{t}_k\rbrace$, for $1\leq k\leq n$ in which the $\textbf{L}_k\in \mathcal{L}_m$ are graph Laplacians to be regressed on covariate vectors $\textbf{t}_k= (t_k^1,\ldots, t_k^u)$, and consider the regression error model 
\begin{equation}\label{eq:error model}
\pi_\nu^{-1}(\text{F}_\alpha(\textbf{L}_k))=\text{vech}^*(\textbf{D}_0+\sum_{w=1}^{u}t_k^w\textbf{D}_w)+\boldsymbol{\epsilon}, \; 
\quad {\boldsymbol{\epsilon}}\sim\mathcal{N}_{m(m-1)/2} (\textbf{0}, \mathbf{\Omega} ) . 
\end{equation}
In general $\mathbf{\Omega}$ has a large number of elements, so in practice it is often necessary to restrict  $\mathbf{\Omega}$ to be diagonal or even isotropic, $\mathbf{\Omega} = \omega^2 \mathbf{I}_{m(m-1)/2}$.

 When using the power Euclidean metric we take $\nu=\textbf{0}$
and the parameters $\lbrace{{\textbf{D}}}_0,\ldots, {{\textbf{D}}}_u\rbrace$  in (\ref{eq:error model}) are $(m-1) \times (m-1)$ symmetric matrices, and
they are estimated by solving
\begin{align}\label{eq:regression model}
(\hat{{\textbf{D}}}_0,\ldots, \hat{{\textbf{D}}}_u)
&=\text{arg}\min_{{{\textbf{D}}}_0,\ldots, {{{\textbf{D}}}}_u}\sum_{k=1}^n
 \| \pi_\nu^{-1}(\text{F}_\alpha(\textbf{L}_k)) - \text{vech}^*(
   \textbf{D}_0+\sum_{w=1}^{u}t_k^w\textbf{D}_w) \|^2 .
\end{align}
The fitted values are
\begin{align}
\textbf{f}(\textbf{t}_k)&=\hat{\textbf{L}}_k=\text{P}_\mathcal{L}\left(\text{G}_\alpha \left(\pi_\nu \left(\text{vech}^*\left(\hat{\textbf{D}}_0+\sum_{w=1}^{u}t_k^w\hat{\textbf{D}}_w\right)\right)\right)\right)\in \mathcal{L}_m,\end{align}
and so $\hat{\textbf{L}}_k$ predicts a graph Laplacian with covariates $\textbf{t}_k$. The optimisation in (\ref{eq:regression model}) is convex and the parameters of the regression line are found using the standard least squares approach in the tangent space. 
This optimisation reduces element-wise for $1\leq i,j \leq m$, to $m(m-1)/2$ independent optimisations.
A similar model can be used for the Procrustes power metric but with $\nu= \hat{\eta} $ and with the $\text{vec}$ operator in place of the $\text{vech*}$ operators.

A test for the significance of covariate $t^w$ involves the hypotheses $H_0:$ $\textbf{D}_w=\textbf{0}$ and $H_1:$ $\textbf{D}_w\neq\textbf{0}$. By Wilks' Theorem \citep{Wilks62}, if $H_0$ is true then the likelihood ratio test statistic is 
\begin{align}\label{eq: regression significance}
T^\ell=-2\log \Delta = -2 \left( \sup_{\mathcal{D},\textbf{D}_w=\textbf{0}} \ell(\mathcal{D})
-\sup_{\mathcal{D},\textbf{D}_w\neq\textbf{0}} \ell(\mathcal{D}) \right)\sim \chi^2_\frac{m(m-1)}{2},
\end{align}
 approximately when $n$ is large, where $\mathcal{D}=\lbrace\textbf{D}_0,\ldots,\textbf{D}_u, \mathbf{\Omega} \rbrace$, $\ell$ is the log-likelihood function under model (\ref{eq:error model}), and 
$\mathbf{\Omega}$ is a diagonal matrix. Using equation (\ref{eq: regression significance}) $H_0$ is rejected in favour of $H_1$ at the $100a\%$ significance level if $T^\ell$ is greater than the $(1-a)$ quantile of $\chi^2_\frac{m(m-1)}{2}$.
 
For the Austen and Dickens data, each novel, represented by a graph Laplacian $\textbf{L}_k$ is paired with the year, $t_k$, the novel was written. We regress the $\lbrace\textbf{L}_k\rbrace$ on the $\lbrace t_k\rbrace$ using the method above with $u=1$ for each author.
To visualise the regression lines in Figure \ref{fig:mds and PC COORDINATES} we find the fitted values $\textbf{f}(t_k)$ for many values of $t_k$ for the specific metrics, and project these to the PC1 and PC2 space.  For each metric the regression lines seem to fit the data well, and could be used to see how writing styles have changed over time.
When the test for regression was performed on the novels the p-values were extremely small ($<10^{-16}$) for both the Austen and Dickens regression lines, for the Euclidean, square root Euclidean and Procrustes size-and-shape metrics. Hence there is very strong evidence to believe that the writing style of both authors changes with time, regardless of which metric we choose. %done this for m=1000

\subsection{A central limit theorem}
Consider independent random samples  $\textbf{A}_{k}, k=1,\ldots,n$ where $F_\alpha(\textbf{A}_k)$ have a distribution with mean $\mathbb{E}( F_\alpha(\textbf{A}))$. 
As the extrinsic mean is based on the arithmetic mean for the power Euclidean metrics, a central limit holds for the sample mean graph Laplacian, under the condition var$(F_\alpha(\textbf{A}))_{ij})$ is finite.
\begin{result}\label{result:central limit} For any power Euclidean metric 
\begin{align*}
\sqrt{n} \; 
{\rm vech}^*\left(
 \hat{\eta} -
 {\eta} \right)
 \xrightarrow[]{D}  \mathcal{N}_\frac{m(m-1)}{2}\big(\textbf{0}, \mathbf{\Sigma} \big),
\end{align*}
as $n \to \infty$, and recall ${\rm vech}^*$ is the {\rm vech} operator but with $\surd 2$ multiplying the terms corresponding to the off-diagonal, and $\mathbf{\Sigma}$ is a finite variance matrix.
\end{result}
When $\alpha=1$ this result is similar to that in \citet{ginestet2017hypothesis} although they work directly in $\mathcal{L}_m$ whereas we work in the embedding space.  For the Procrustes power metric a similar central limit theorem result follows providing the more stringent conditions of \citet{bhattacharya2005} hold. %, the validity of these conditions may be worth exploring in the future.

\subsection{Hypothesis tests} \label{sec:two sample test}
Consider two populations $\mathcal{A}$ and $\mathcal{B}$ of $m \times m$ graph Laplacians with corresponding population means $\mu_A$ and $\mu_B$ defined in (\ref{eq: pop mean}) . Given two independent random samples $\lbrace \textbf{A}_{1}, \textbf{A}_{2},\ldots, \textbf{A}_{n_A} \rbrace$ and $\lbrace \textbf{B}_{1}, \textbf{B}_{2},\ldots, \textbf{B}_{n_B} \rbrace$ respectively from $\mathcal{A}$ and $\mathcal{B}$, the goal is to test the hypotheses
\begin{align*}
&H_{0}: \mu_A=\mu_B \text{ and } H_{1} : \mu_A\neq \mu_B. 
\end{align*} 
We define the test statistic as
$ T=d(\hat{\textbf{A}}, \hat{\textbf{B}})^{2}$,
 where $\hat{\textbf{A}}$ and $\hat{\textbf{B}}$ are defined by $\hat{\eta}$ in (\ref{eq:sample mean}) for the sets $\mathcal{A}$ and $\mathcal{B}$ respectively and using a suitable metric.  Any Euclidean or Procrustes power metric is suitable to use, we however will just consider the Euclidean $T_E=d_1(\hat{\textbf{A}}_E,  \hat{\textbf{B}}_E)^{2}$; the square root Euclidean $T_{H}={d}_\frac{1}{2}(\hat{\textbf{A}}_H,  \hat{\textbf{B}}_H)^{2}$; and the Procrustes size-and-shape 
 $T_S={d}_{\frac{1}{2},S}(\hat{\textbf{A}}_S, \hat{\textbf{B}}_S)^{2}$, where the subscripts $\{E,H,S\}$ refer to whether the Euclidean, square root or Procrustes size-and-shape means have been used, respectively.

Using Result \ref{result:central limit},  the distribution of the test statistics for large $n$ is given as follows for the power Euclidean metric. 
\begin{result}\label{result: central limit theorem euclidean}
Consider independent random samples of networks of size $n_A$ and $n_B$. For the power Euclidean metric under the null hypothesis, $H_0$: $\mu_A=\mu_B$, as $n_A, n_B \to \infty$, such that $n_A/n_B \to r \in (0,\infty)$: 
\begin{align}\label{eq:asym1}
 \frac{n_An_B}{n_A + n_B} d_\alpha( \hat{\textbf{A}} , \hat{\textbf{B}} )^{2}  \; \xrightarrow[]{D} \;    \sum_{i=1}^{m(m-1)/2}\delta_{i} \chi_{1}^{2} ,
\end{align} 
in which each $\chi_{1}^{2}$ is independent and $\delta_{i}$ are the $m(m-1)/2$ non-zero eigenvalues of $\mathbf{\Sigma}=\frac{n_B\boldsymbol{\Sigma}_A+n_A\boldsymbol{\Sigma}_B}{n_A+n_B}$, where $\boldsymbol{\Sigma}_A$ and $\boldsymbol{\Sigma}_B$ are the population covariance matrices 
from Result \ref{result:central limit} for the respective samples.
\end{result}

In practice $\mathbf{\Sigma}$ needs to be estimated, which can be very high dimensional. In our application with $m=1000$ this is a symmetric matrix with $M(M+1)/2$ parameters where $M=m(m-1)/2 = 499500$.  
One approach is to use the shrinkage estimator from \citet{schafer2005shrinkage}, as employed by 
\citet{ginestet2017hypothesis}, but this is impractical for our application with $m=1000$. If we assume a diagonal matrix $\mathbf{\Sigma} = \mathbf{\Lambda^*}$ then 
the $\delta_i$ correspond to the variances of individual components of the difference in means, and these can be estimated consistently from method of moments estimators.  
A further very simple model would be to have an isotropic covariance matrix with covariance matrix $\mathbf{\Sigma} = \sigma^2 \textbf{I}_{m(m-1)/2}$, which requires estimation of a single 
variance parameter $\sigma^2$.  Note that the likelihood ratio test for regression with test statistic $-2\log \Delta$ in Section \ref{sec:regression} gives an alternative test for equality of means when the covariates are group labels, 
but the additional assumption of normality for the observations needs to be made in that case.   

An alternative non-parametric test, which does not depend on large sample asymptotics is a random permutation test, similar to \citet{10.2307/41000409} as follows.
\begin{algorithm}[H]
  \begin{algorithmic}[1]
    \STATE Calculate the test statistics between $\mathcal{A}$ and $\mathcal{B}$, given by $T=T(\mathcal{A}, \mathcal{B})$ .
   \STATE Generate random sets $\mathcal{A}^*$ and $\mathcal{B}^*$ of size $\vert\mathcal{A}\vert$ and $\vert\mathcal{B}\vert$ respectively, by randomly sampling without replacement from $\mathcal{A}\cup\mathcal{B}$.
    \STATE Compute the test statistic of sets $\mathcal{A}^*$ and $\mathcal{B}^*$, given by $T^*=T(\mathcal{A}^*, \mathcal{B}^*)$.
    \STATE Repeat steps 2 and 3 $r$ times, to give test statistics $T^*_1, T^*_2, \ldots, T^*_r$ .
    \STATE Order the test statistics $T^*(1)\leq T^*(2)\leq \ldots \leq T^*(r)$.
    \STATE Calculate the p-value, which is $1-\frac{j}{r}$  for the minimum $1 \leq j\leq r-1$ satisfying \newline $T^*(j)< T\leq T^*(j+1)$, unless $T\leq T^*(1)$, in which case the p-value is 1 or if $T> T^*(r)$, in which case the p-value is 0.
  \end{algorithmic}
  \caption{Random permutation test to test the equality of means for two sets of graph Laplacians, $\mathcal{A}$ and $\mathcal{B}$, using the test statistic $T$.}\label{algorithm: perm}
\end{algorithm} 

A limitation of using the permutation test is it assumes exchangeability of the observations under the null hypothesis \citep{doi:10.1198/016214506000001400}. This means under the null hypothesis the populations $\mathcal{A}$ and $\mathcal{B}$ are assumed identical. A test based on the bootstrap is an alternative possibility, which requires weaker assumptions about $\mathcal{A}$ and $\mathcal{B}$, see for example \citet{doi:10.1198/016214506000001400}. 

For the Austen and Dickens data have test statistics $T_E = 0.0011$, $T_{H}=0.2759$, $T_S=0.0691$.  We compute the p-value from the permutation test with $r=199$ permutations for 
each of $T_E, T_{H}, T_S$ and in each case all permuted values were less than the observed statistics for the data. Hence, in each 
case the estimated p-value is zero, indicating very strong evidence for a difference in mean graph Laplacian. 

\subsection{Exploring differences between authors}\label{explore}
The result that the Austen and Dickens novels are highly significantly different is not unexpected due to the 
clear difference between Austen and Dickens' novels seen in the PCA plot in Figure \ref{fig:mds and PC COORDINATES}. Also, high-dimensional 
multivariate tests of global differences are often significant due to the nature  of high-dimensional spaces, where random observations become approximately orthogonal to each other as the dimension increases \citep{Halletal05}. To address this issue further we now focus on the main edge-specific differences between the Austen and Dickens 
networks, which is of strong practical interest. 

In particular we examine the off-diagonal elements of   $\hat{\mu}^{Austen}_E-\hat{\mu}^{Dickens}_E$ 
%$$ P_{\cal L}( \textbf{G}_\alpha(\hat \eta_{Dickens})) -  P_{\cal L}( \textbf{G}_\alpha( \hat \eta_{Austen})) , $$ 
i.e. the differences in the mean weighted adjacency matrix, and compare them to appropriate measures of 
standard error of the differences using a $z$-statistic. 
The histograms of the off-diagonal individual graph Laplacians are heavy tailed, and a plot of sample standard deviations versus sample means 
show an overall average linear increase with approximate slope $\beta = 0.2$, but with a large spread. We shall use this relationship in a regularised estimate of 
our choice of standard error. 

For a particular co-occurrence pair of words we have weighted adjacency values $x_i, i=1,\ldots,n_A$ and $y_j, j=1,\ldots,n_B$ with sample means
$\bar x$ and $\bar y$, and sample standard deviations $s_x$ and $s_y$. For our analysis here we use the Euclidean mean graph Laplacians. 
We estimate the variance in our sample with a weighted average of the sample variance and an estimate based on the linear relationship 
between the mean and standard deviation, and in particular  the population pooled variance is estimated by 
$$s_p^2 = \frac{ {n_A}( w_{A} s_x^2 + (1-w_{A})\beta^2 \bar x^2) + {n_B}(w_B s_y^2 + (1-w_B) \beta^2 \bar y^2) }{ (n_A+n_B - 2) }, $$
where the weights are taken as $w_A = n_A/N, w_B=n_B/N$, where we take $N=200$.  Note that if all values in one of the samples are $0$ (due to no word co-occurrence pairings in
any of that author's books) then we drop that word pairing from further analysis, as we are only interested in the relative usage of the word occurrences that are 
actually used by both authors. 
A univariate $z$-statistic for comparing adjacencies is then 
\begin{equation}
z = \frac{ \bar x - \bar y }{ (q + s_p) \sqrt{\frac{1}{n_A} + \frac{1}{n_B} } } ,  \label{zstat}
\end{equation}
where we include the regularizing offset $q>0$ to avoid highlighting very small differences in mean adjacency with very small standard errors.  The value for $q$ is chosen as 
the median of all $s_p$ values under consideration. 
%For the most negative 100 values of $z$ $(z  < -13.57)$ we display a network in the upper plots of Figure \ref{Top100Pairs} 
%with an edge between each of these 100 word pairings, which 
%gives the most prominent pairs of words which co-occur more in Austen than Dickens. Recall words co-occur in a pair together if they lie within a span of 5 words to the left or tight.  For the top 100 largest values of $z$ $(z > 10.12)$ we display a network in the lower plot of Figure \ref{Top100Pairs}
%with an edge between each of these 100 word pairings, which 
%gives the most prominent pairs of words which co-occur more in Dickens than Austen. 

\begin{figure}[htbp]
\begin{center}
\includegraphics[width=9cm,height=9cm]{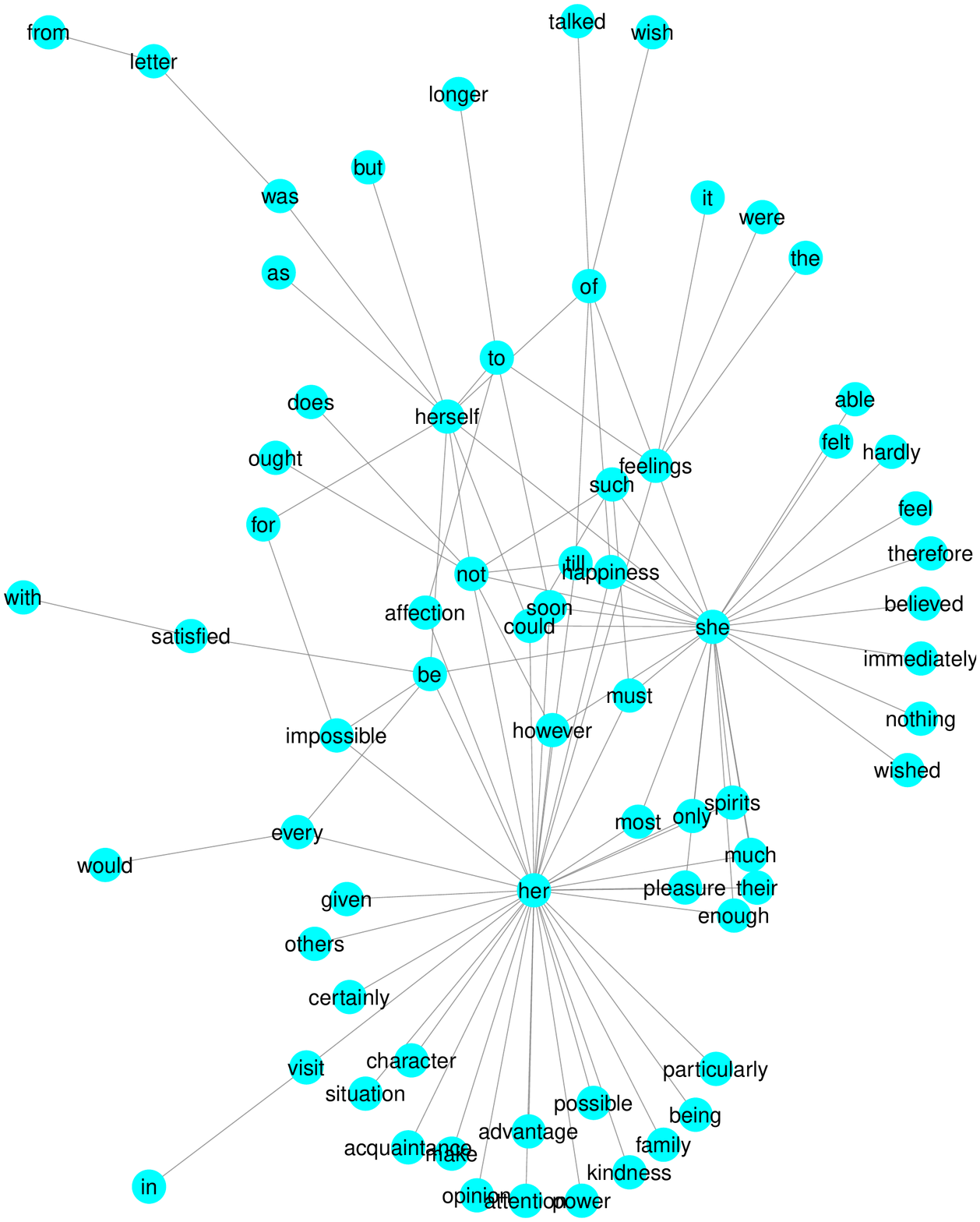}
\includegraphics[width=9cm,height=9cm]{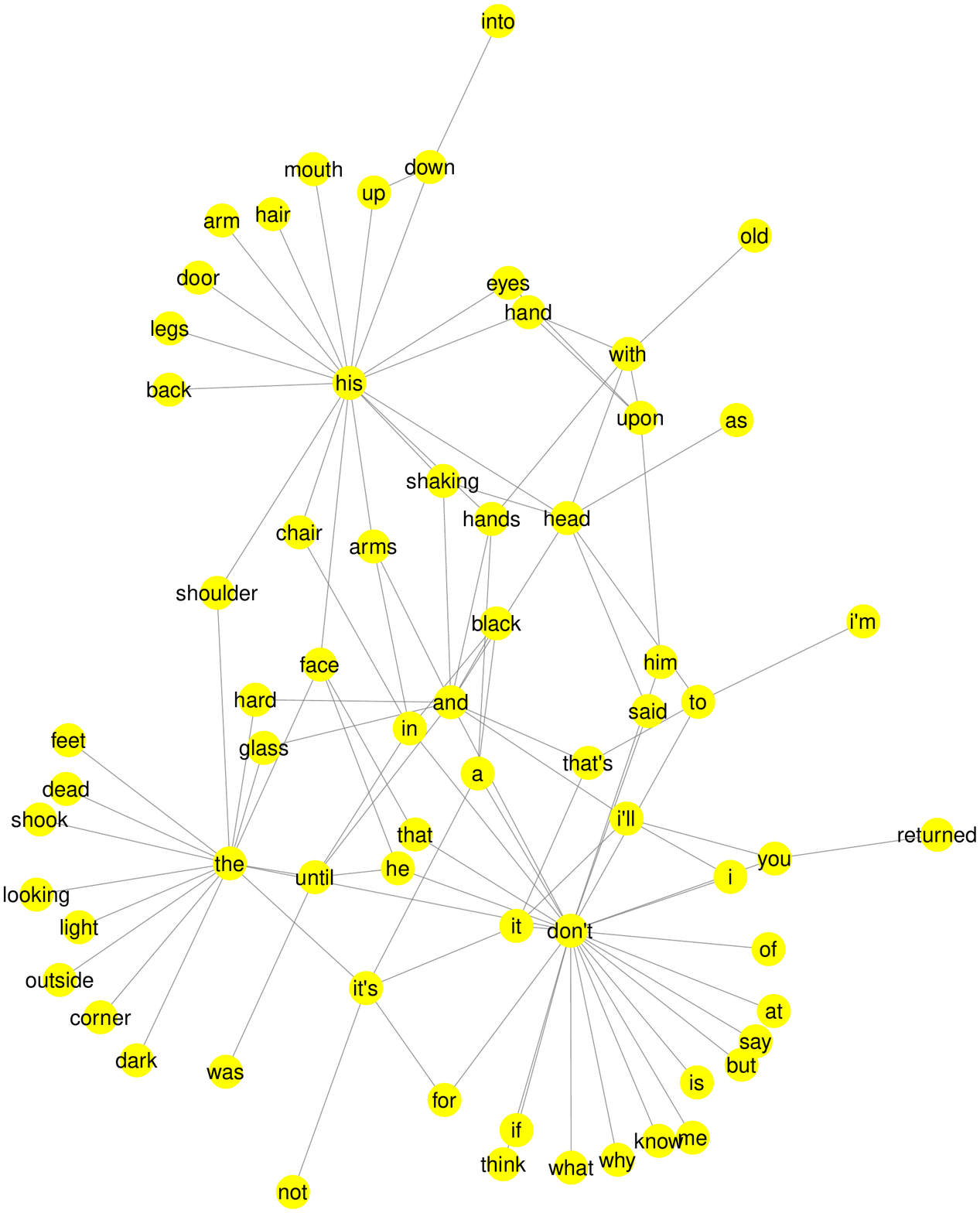}
\end{center}
 \caption{\footnotesize{\textit{Networks displaying the top 100 pairs of words ranked according to the $z$-statistic in (\ref{zstat}), with more prominent co-occurrences used by Austen (top) and the more prominent co-occurrences used by Dickens
 (below). }}}\label{Top100Pairs}
\end{figure}

Exploratory graphical displays are given in Figure \ref{Top100Pairs} as networks 
where an edge is drawn between two words if they appear in the 
top 100 pairs of words ranked according to the $z$-statistic in (\ref{zstat}). The plots show the 
more prominent co-occurrences used by Austen and the more prominent co-occurrences used by Dickens, respectively. The displays 
illuminate striking differences between the novelists. For Austen there are very common pairings of words with 
{\it her}, {\it she}, {\it herself}, which form important hubs in this network. Austen also pairs these hubs with more emotional words 
{\it feelings}, {\it felt}, {\it feel}, {\it kindness}, {\it happiness}, {\it affection}, 
{\it pleasure} and stronger words  {\it power}, {\it attention}, {\it must}, {\it certainly}, {\it advantage} and {\it opinion}. Also we see more use of {\it letter} in Austen, which is a literary 
device often used by the author. 
For Dickens there are more common uses of 
abbreviations, especially {\it don't} which is an important hub, and also {\it it's}, {\it i'll} and {\it that's}. In contrast the Austen network highlights {\it not}.  Dickens also more 
prominently pairs body parts {\it arm, arms, eyes, feet, hair, hand, hands, head, mouth, face, shoulder, legs} in combination with the strong hubs {
\it his} and {\it the}. These hubs are also paired with other objects, such as {\it door, chair, glass}. Finally, Dickens has the 
more prominent use of pairs with a sombre word, such as {\it dark}, {\it black} and {\it dead}, which might have been expected.

\section{Conclusion}\label{sec:conclusion}

We have developed a general framework for the statistical analysis of graph Laplacians and considered in particular the power Euclidean, $d_\alpha$, and Procrustes size-and-shape, $d_{\alpha,S}$, metrics. { The framework is extrinsic except when $d_1$ is used.}
Other metrics fit in our extrinsic framework and could be considered.
{One example is the log metric used in \citet{bakker2018dynamic} which uses the embedding $\text{F}_{\log}(\textbf{L})=\sum_{i=1}^l\log{(\xi_i)}\textbf{u}_i\textbf{u}_i^T$, and $\text{F}_{\log}(\textbf{L})=\lim_{\alpha\rightarrow 0}\frac{1}{\alpha}(\text{F}_\alpha(\textbf{L})-\text{F}_0(\textbf{L})$) where we define $0^0=0$ in $\text{F}_0$ and $l$ is the rank of $\textbf{L}$. The metric is then $d_{\log}(\textbf{L}_1, \textbf{L}_2)=\Vert\text{F}_{\log}(\textbf{L}_1)-\text{F}_{\log}(\textbf{L}_2)\Vert$.}  {The log embedding is a natural embedding to consider in more detail in further work and to compare the properties of the log embedding methods with using $\textbf{F}_\alpha$. The log embedding 
been used successfully in many applications, for example for symmetric positive definite matrices extracted from diffusion tensor imaging (DTI) in \citet{bhattacharya2017omnibus}. An extrinsic regression model similar to the ours but using kernel regression has been developed for manifolds by \citet{lin2017extrinsic},  and in future work it would be worth investigating if their results extend to the $\textbf{F}_\alpha$  embedding. }

 Another metric to consider is the element-wise metric of the form\\$d_{\rho}^*(\textbf{L}_{1},\textbf{L}_{2}) = (\sum_i\sum_j\vert(\textbf{L}_{1})_{ij}-(\textbf{L}_{2})_{ij}\vert^{\rho})^\frac{1}{\rho}$. Of particular interest would be comparing 
 $\rho=2$, which is the Frobenius/Euclidean norm $d_1$, with $\rho=1$ which can be similar to the square root norm (and is identical for diagonal matrices).

One practical issue is the estimation of covariance matrices in models for graph Laplacians. 
In general for $m$ by $m$ graph Laplacians the covariance matrix $\mathbf{\Omega}$ has a very large number $\frac{m(m-1)(m^2-m+2)}{8}$ of parameters in the regression model (\ref{eq:error model}). A very 
large number $n$ of networks would be needed for estimating the most general 
form of the model, and so in practice we assume that the covariance matrix is diagonal. 
Extending to non-diagonal parameterised covariance structures could also be 
sometimes feasible, e.g. autoregressive models or covariance structure based on the spatial location of the nodes if that makes sense in particular applications. 

In our application we have ordered the 1000 word lists, and we used the ordering of the most common words from 
the combined set of novels of both authors. Although the ordering of most common words is different 
in each novel, there is consistency in the broad ordering of common words. It does make sense 
in our application to keep a common ordering, and the effect of using the Procrustes distance versus the power metric 
is not so large, as we have seen. However, in other applications where there is a less obvious correspondence 
between nodes, the Procrustes distance could be very different. 

Our methodology gives appropriate results for comparing co-occurrence networks for Jane Austen and Charles Dickens novels, but the methodology is widely applicable, for example to neuroimaging networks and social networks, 
and such applications will be explored in further work.

\bibliography{paperbib}

\begin{thebibliography}{40}
% BibTex style file: imsart-nameyear.bst, 2017-11-03
% Default style options (sort=1,type=nameyear).
% Used options (sort=1,type=nameyear).

\bibitem[\protect\citeauthoryear{Amaral, Dryden and
  Wood}{2007}]{doi:10.1198/016214506000001400}
\begin{barticle}[author]
\bauthor{\bsnm{Amaral},~\bfnm{G.~J.~A}\binits{G.~J.~A.}},
  \bauthor{\bsnm{Dryden},~\bfnm{I.~L}\binits{I.~L.}} \AND
  \bauthor{\bsnm{Wood},~\bfnm{Andrew T.~A}\binits{A.~T.~A.}}
(\byear{2007}).
\btitle{Pivotal Bootstrap Methods for k-Sample Problems in Directional
  Statistics and Shape Analysis}.
\bjournal{Journal of the American Statistical Association}
\bvolume{102}
\bpages{695-707}.
\bdoi{10.1198/016214506000001400}
\end{barticle}
\endbibitem

\bibitem[\protect\citeauthoryear{Anderson}{2018}]{rosqp18}
\begin{bmanual}[author]
\bauthor{\bsnm{Anderson},~\bfnm{Eric}\binits{E.}}
(\byear{2018}).
\btitle{rosqp: Quadratic Programming Solver using the OSQP Library}
\bnote{R package version 0.1.0}.
\end{bmanual}
\endbibitem

\bibitem[\protect\citeauthoryear{Bakker, Halappanavar and
  Sathanur}{2018}]{bakker2018dynamic}
\begin{barticle}[author]
\bauthor{\bsnm{Bakker},~\bfnm{Craig}\binits{C.}},
  \bauthor{\bsnm{Halappanavar},~\bfnm{Mahantesh}\binits{M.}} \AND
  \bauthor{\bsnm{Sathanur},~\bfnm{Arun~Visweswara}\binits{A.~V.}}
(\byear{2018}).
\btitle{Dynamic graphs, community detection, and {R}iemannian geometry}.
\bjournal{Applied Network Science}
\bvolume{3}
\bpages{3}.
\end{barticle}
\endbibitem

\bibitem[\protect\citeauthoryear{Bhattacharya and
  Lin}{2017}]{bhattacharya2017omnibus}
\begin{barticle}[author]
\bauthor{\bsnm{Bhattacharya},~\bfnm{Rabi}\binits{R.}} \AND
  \bauthor{\bsnm{Lin},~\bfnm{Lizhen}\binits{L.}}
(\byear{2017}).
\btitle{Omnibus CLTs for Fr{\'e}chet means and nonparametric inference on
  non-Euclidean spaces}.
\bjournal{Proceedings of the American Mathematical Society}
\bvolume{145}
\bpages{413--428}.
\end{barticle}
\endbibitem

\bibitem[\protect\citeauthoryear{Bhattacharya and
  Patrangenaru}{2003}]{bhattacharya2003}
\begin{barticle}[author]
\bauthor{\bsnm{Bhattacharya},~\bfnm{Rabi}\binits{R.}} \AND
  \bauthor{\bsnm{Patrangenaru},~\bfnm{Vic}\binits{V.}}
(\byear{2003}).
\btitle{Large sample theory of intrinsic and extrinsic sample means on
  manifolds}.
\bjournal{Ann. Statist.}
\bvolume{31}
\bpages{1--29}.
\bdoi{10.1214/aos/1046294456}
\end{barticle}
\endbibitem

\bibitem[\protect\citeauthoryear{Bhattacharya and
  Patrangenaru}{2005}]{bhattacharya2005}
\begin{barticle}[author]
\bauthor{\bsnm{Bhattacharya},~\bfnm{Rabi}\binits{R.}} \AND
  \bauthor{\bsnm{Patrangenaru},~\bfnm{Vic}\binits{V.}}
(\byear{2005}).
\btitle{Large sample theory of intrinsic and extrinsic sample means on
  manifolds—II}.
\bjournal{Ann. Statist.}
\bvolume{33}
\bpages{1225--1259}.
\bdoi{10.1214/009053605000000093}
\end{barticle}
\endbibitem

\bibitem[\protect\citeauthoryear{Cootes et~al.}{1994}]{Cootetal94}
\begin{bincollection}[author]
\bauthor{\bsnm{Cootes},~\bfnm{T.~F.}\binits{T.~F.}},
  \bauthor{\bsnm{Taylor},~\bfnm{C.~J.}\binits{C.~J.}},
  \bauthor{\bsnm{Cooper},~\bfnm{D.~H.}\binits{D.~H.}} \AND
  \bauthor{\bsnm{Graham},~\bfnm{J.}\binits{J.}}
(\byear{1994}).
\btitle{Image search using flexible shape models generated from sets of
  examples}.
In \bbooktitle{Statistics and Images: Vol. 2}
(\beditor{\bfnm{K.~V.}\binits{K.~V.}~\bsnm{Mardia}}, ed.)
\bpages{111-139}.
\bpublisher{Carfax}, \baddress{Oxford}.
\end{bincollection}
\endbibitem

\bibitem[\protect\citeauthoryear{De~Klerk}{2006}]{de2006aspects}
\begin{bbook}[author]
\bauthor{\bsnm{De~Klerk},~\bfnm{Etienne}\binits{E.}}
(\byear{2006}).
\btitle{Aspects of semidefinite programming: interior point algorithms and
  selected applications}
\bvolume{65}.
\bpublisher{Springer Science \& Business Media}.
\end{bbook}
\endbibitem

\bibitem[\protect\citeauthoryear{Dryden}{2019}]{Dryden-shapes}
\begin{bmanual}[author]
\bauthor{\bsnm{Dryden},~\bfnm{I.~L.}\binits{I.~L.}}
(\byear{2019}).
\btitle{{\tt shapes} package}
\bpublisher{R Foundation for Statistical Computing},
\baddress{Vienna, Austria}
\bnote{Contributed package, Version 1.2.5.}
\end{bmanual}
\endbibitem

\bibitem[\protect\citeauthoryear{Dryden, Koloydenko and
  Zhou}{2009}]{10.2307/30242879}
\begin{barticle}[author]
\bauthor{\bsnm{Dryden},~\bfnm{Ian~L.}\binits{I.~L.}},
  \bauthor{\bsnm{Koloydenko},~\bfnm{Alexey}\binits{A.}} \AND
  \bauthor{\bsnm{Zhou},~\bfnm{Diwei}\binits{D.}}
(\byear{2009}).
\btitle{Non-{Euclidean} Statistics for Covariance Matrices, with Applications
  to Diffusion Tensor Imaging}.
\bjournal{The Annals of Applied Statistics}
\bvolume{3}
\bpages{1102-1123}.
\end{barticle}
\endbibitem

\bibitem[\protect\citeauthoryear{Dryden and Mardia}{2016}]{MR3559734}
\begin{bbook}[author]
\bauthor{\bsnm{Dryden},~\bfnm{Ian~L.}\binits{I.~L.}} \AND
  \bauthor{\bsnm{Mardia},~\bfnm{Kanti~V.}\binits{K.~V.}}
(\byear{2016}).
\btitle{Statistical shape analysis with applications in {R}},
\bedition{second} ed.
\bseries{Wiley Series in Probability and Statistics}.
\bpublisher{John Wiley \& Sons, Ltd., Chichester}.
\bdoi{10.1002/9781119072492}
\bmrnumber{3559734}
\end{bbook}
\endbibitem

\bibitem[\protect\citeauthoryear{{Dryden}, {Pennec} and
  {Peyrat}}{2010}]{Drydenetal10}
\begin{barticle}[author]
\bauthor{\bsnm{{Dryden}},~\bfnm{Ian~L.}\binits{I.~L.}},
  \bauthor{\bsnm{{Pennec}},~\bfnm{Xavier}\binits{X.}} \AND
  \bauthor{\bsnm{{Peyrat}},~\bfnm{Jean-Marc}\binits{J.-M.}}
(\byear{2010}).
\btitle{{Power Euclidean metrics for covariance matrices with application to
  diffusion tensor imaging}}.
\bjournal{arXiv e-prints}
\bpages{arXiv:1009.3045}.
\end{barticle}
\endbibitem

\bibitem[\protect\citeauthoryear{Evert}{2008}]{evert2008corpora}
\begin{barticle}[author]
\bauthor{\bsnm{Evert},~\bfnm{Stefan}\binits{S.}}
(\byear{2008}).
\btitle{Corpora and collocations}.
\bjournal{Corpus linguistics. An international handbook}
\bvolume{2}
\bpages{1212--1248}.
\end{barticle}
\endbibitem

\bibitem[\protect\citeauthoryear{Fletcher et~al.}{2004}]{fletcher2004principal}
\begin{barticle}[author]
\bauthor{\bsnm{Fletcher},~\bfnm{P~Thomas}\binits{P.~T.}},
  \bauthor{\bsnm{Lu},~\bfnm{Conglin}\binits{C.}},
  \bauthor{\bsnm{Pizer},~\bfnm{Stephen~M}\binits{S.~M.}} \AND
  \bauthor{\bsnm{Joshi},~\bfnm{Sarang}\binits{S.}}
(\byear{2004}).
\btitle{Principal geodesic analysis for the study of nonlinear statistics of
  shape}.
\bjournal{IEEE transactions on medical imaging}
\bvolume{23}
\bpages{995--1005}.
\end{barticle}
\endbibitem

\bibitem[\protect\citeauthoryear{Fr{\'e}chet}{1948}]{Frechet1948}
\begin{barticle}[author]
\bauthor{\bsnm{Fr{\'e}chet},~\bfnm{Maurice}\binits{M.}}
(\byear{1948}).
\btitle{Les {\'e}l{\'e}ments al{\'e}atoires de nature quelconque dans un espace
  distanci{\'e}}.
\bjournal{Annales de l'institut Henri Poincaré}
\bvolume{10}
\bpages{215-310}.
\end{barticle}
\endbibitem

\bibitem[\protect\citeauthoryear{Fu et~al.}{2020}]{CVXR18}
\begin{bmanual}[author]
\bauthor{\bsnm{Fu},~\bfnm{Anqi}\binits{A.}},
  \bauthor{\bsnm{Narasimhan},~\bfnm{Balasubramanian}\binits{B.}},
  \bauthor{\bsnm{Kang},~\bfnm{David~W}\binits{D.~W.}},
  \bauthor{\bsnm{Diamond},~\bfnm{Steven}\binits{S.}} \AND
  \bauthor{\bsnm{Miller},~\bfnm{John}\binits{J.}}
(\byear{2020}).
\btitle{CVXR: Disciplined Convex Optimization}
\bnote{R package version 1.0-1}.
\end{bmanual}
\endbibitem

\bibitem[\protect\citeauthoryear{Ginestet
  et~al.}{2017}]{ginestet2017hypothesis}
\begin{barticle}[author]
\bauthor{\bsnm{Ginestet},~\bfnm{Cedric~E}\binits{C.~E.}},
  \bauthor{\bsnm{Li},~\bfnm{Jun}\binits{J.}},
  \bauthor{\bsnm{Balachandran},~\bfnm{Prakash}\binits{P.}},
  \bauthor{\bsnm{Rosenberg},~\bfnm{Steven}\binits{S.}},
  \bauthor{\bsnm{Kolaczyk},~\bfnm{Eric~D}\binits{E.~D.}} \betal{et~al.}
(\byear{2017}).
\btitle{Hypothesis testing for network data in functional neuroimaging}.
\bjournal{The Annals of Applied Statistics}
\bvolume{11}
\bpages{725--750}.
\end{barticle}
\endbibitem

\bibitem[\protect\citeauthoryear{Hall, Marron and Neeman}{2005}]{Halletal05}
\begin{barticle}[author]
\bauthor{\bsnm{Hall},~\bfnm{Peter}\binits{P.}},
  \bauthor{\bsnm{Marron},~\bfnm{J.~S.}\binits{J.~S.}} \AND
  \bauthor{\bsnm{Neeman},~\bfnm{Amnon}\binits{A.}}
(\byear{2005}).
\btitle{Geometric representation of high dimension, low sample size data}.
\bjournal{J. R. Stat. Soc. Ser. B Stat. Methodol.}
\bvolume{67}
\bpages{427--444}.
\bmrnumber{MR2155347}
\end{barticle}
\endbibitem

\bibitem[\protect\citeauthoryear{Higham}{1988}]{higham1988computing}
\begin{barticle}[author]
\bauthor{\bsnm{Higham},~\bfnm{Nicholas~J}\binits{N.~J.}}
(\byear{1988}).
\btitle{Computing a nearest symmetric positive semidefinite matrix}.
\bjournal{Linear algebra and its applications}
\bvolume{103}
\bpages{103--118}.
\end{barticle}
\endbibitem

\bibitem[\protect\citeauthoryear{{Charles Dickens Info}}{2020}]{charlesDickens}
\begin{bmisc}[author]
\bauthor{\bsnm{{Charles Dickens Info}}}
(\byear{2020}).
\btitle{Charles Dickens Timeline}.
\bnote{\url{https://www.charlesdickensinfo.com/life/timeline/}, Last accessed
  on 2020-08-17}.
\end{bmisc}
\endbibitem

\bibitem[\protect\citeauthoryear{Kendall et~al.}{1999}]{Kendetal99}
\begin{bbook}[author]
\bauthor{\bsnm{Kendall},~\bfnm{D.~G.}\binits{D.~G.}},
  \bauthor{\bsnm{Barden},~\bfnm{D.}\binits{D.}},
  \bauthor{\bsnm{Carne},~\bfnm{T.~K.}\binits{T.~K.}} \AND
  \bauthor{\bsnm{Le},~\bfnm{H.}\binits{H.}}
(\byear{1999}).
\btitle{Shape and Shape Theory}.
\bpublisher{Wiley}, \baddress{Chichester}.
\end{bbook}
\endbibitem

\bibitem[\protect\citeauthoryear{Kent}{1994}]{kent1994complex}
\begin{barticle}[author]
\bauthor{\bsnm{Kent},~\bfnm{John~T}\binits{J.~T.}}
(\byear{1994}).
\btitle{The complex Bingham distribution and shape analysis}.
\bjournal{Journal of the Royal Statistical Society. Series B (Methodological)}
\bpages{285--299}.
\end{barticle}
\endbibitem

\bibitem[\protect\citeauthoryear{Kolaczyk}{2009}]{kolaczyk2009statistical}
\begin{bbook}[author]
\bauthor{\bsnm{Kolaczyk},~\bfnm{Eric~D}\binits{E.~D.}}
(\byear{2009}).
\btitle{Statistical analysis of network data: methods and models}.
\bpublisher{Springer Science \& Business Media}.
\end{bbook}
\endbibitem

\bibitem[\protect\citeauthoryear{Kolaczyk et~al.}{2020}]{kolaczyk2020}
\begin{barticle}[author]
\bauthor{\bsnm{Kolaczyk},~\bfnm{Eric~D.}\binits{E.~D.}},
  \bauthor{\bsnm{Lin},~\bfnm{Lizhen}\binits{L.}},
  \bauthor{\bsnm{Rosenberg},~\bfnm{Steven}\binits{S.}},
  \bauthor{\bsnm{Walters},~\bfnm{Jackson}\binits{J.}} \AND
  \bauthor{\bsnm{Xu},~\bfnm{Jie}\binits{J.}}
(\byear{2020}).
\btitle{Averages of unlabeled networks: Geometric characterization and
  asymptotic behavior}.
\bjournal{Ann. Statist.}
\bvolume{48}
\bpages{514--538}.
\bdoi{10.1214/19-AOS1820}
\end{barticle}
\endbibitem

\bibitem[\protect\citeauthoryear{Le}{1995}]{10.2307/1428094}
\begin{barticle}[author]
\bauthor{\bsnm{Le},~\bfnm{Huiling}\binits{H.}}
(\byear{1995}).
\btitle{Mean Size-and-Shapes and Mean Shapes: A Geometric Point of View}.
\bjournal{Advances in Applied Probability}
\bvolume{27}
\bpages{44--55}.
\end{barticle}
\endbibitem

\bibitem[\protect\citeauthoryear{Lin et~al.}{2017}]{lin2017extrinsic}
\begin{barticle}[author]
\bauthor{\bsnm{Lin},~\bfnm{Lizhen}\binits{L.}},
  \bauthor{\bsnm{St.~Thomas},~\bfnm{Brian}\binits{B.}},
  \bauthor{\bsnm{Zhu},~\bfnm{Hongtu}\binits{H.}} \AND
  \bauthor{\bsnm{Dunson},~\bfnm{David~B}\binits{D.~B.}}
(\byear{2017}).
\btitle{Extrinsic local regression on manifold-valued data}.
\bjournal{Journal of the American Statistical Association}
\bvolume{112}
\bpages{1261--1273}.
\end{barticle}
\endbibitem

\bibitem[\protect\citeauthoryear{Mahlberg
  et~al.}{2016}]{doi:10.3366/cor.2016.0102}
\begin{barticle}[author]
\bauthor{\bsnm{Mahlberg},~\bfnm{Michaela}\binits{M.}},
  \bauthor{\bsnm{Stockwell},~\bfnm{Peter}\binits{P.}}, \bauthor{\bparticle{de}
  \bsnm{Joode},~\bfnm{Johan}\binits{J.}},
  \bauthor{\bsnm{Smith},~\bfnm{Catherine}\binits{C.}} \AND
  \bauthor{\bsnm{O'Donnell},~\bfnm{Matthew~Brook}\binits{M.~B.}}
(\byear{2016}).
\btitle{{CLiC Dickens}: novel uses of concordances for the integration of
  corpus stylistics and cognitive poetics}.
\bjournal{Corpora}
\bvolume{11}
\bpages{433-463}.
\bdoi{10.3366/cor.2016.0102}
\end{barticle}
\endbibitem

\bibitem[\protect\citeauthoryear{Masarotto, Panaretos and
  Zemel}{2019}]{Masarottoetal19}
\begin{barticle}[author]
\bauthor{\bsnm{Masarotto},~\bfnm{V.}\binits{V.}},
  \bauthor{\bsnm{Panaretos},~\bfnm{V.~M.}\binits{V.~M.}} \AND
  \bauthor{\bsnm{Zemel},~\bfnm{Y.}\binits{Y.}}
(\byear{2019}).
\btitle{Procrustes Metrics on Covariance Operators and Optimal Transportation
  of Gaussian Processes}.
\bjournal{Sankhya A}
\bvolume{81}
\bpages{172--213}.
\end{barticle}
\endbibitem

\bibitem[\protect\citeauthoryear{{The Jane Austen Society of North
  America}}{2020}]{janeAus}
\begin{bmisc}[author]
\bauthor{\bsnm{{The Jane Austen Society of North America}}}
(\byear{2020}).
\btitle{Jane Austen's Works}.
\bnote{\url{http://jasna.org/austen/works/}, Last accessed on 2020-08-17}.
\end{bmisc}
\endbibitem

\bibitem[\protect\citeauthoryear{Phillips}{1983}]{phillips1983lexical}
\begin{bbook}[author]
\bauthor{\bsnm{Phillips},~\bfnm{M.~K.}\binits{M.~K.}}
(\byear{1983}).
\btitle{Lexical Macrostructure in Science Text}.
\bpublisher{University of Birmingham}.
\end{bbook}
\endbibitem

\bibitem[\protect\citeauthoryear{Pigoli et~al.}{2014}]{Pigolietal14}
\begin{barticle}[author]
\bauthor{\bsnm{Pigoli},~\bfnm{Davide}\binits{D.}},
  \bauthor{\bsnm{Aston},~\bfnm{John A.~D.}\binits{J.~A.~D.}},
  \bauthor{\bsnm{Dryden},~\bfnm{Ian~L.}\binits{I.~L.}} \AND
  \bauthor{\bsnm{Secchi},~\bfnm{Piercesare}\binits{P.}}
(\byear{2014}).
\btitle{Distances and inference for covariance operators}.
\bjournal{Biometrika}
\bvolume{101}
\bpages{409--422}.
\bdoi{10.1093/biomet/asu008}
\bmrnumber{3215356}
\end{barticle}
\endbibitem

\bibitem[\protect\citeauthoryear{Preston and Wood}{2010}]{10.2307/41000409}
\begin{barticle}[author]
\bauthor{\bsnm{Preston},~\bfnm{S.~P.}\binits{S.~P.}} \AND
  \bauthor{\bsnm{Wood},~\bfnm{A.~T.~A.}\binits{A.~T.~A.}}
(\byear{2010}).
\btitle{Two-Sample Bootstrap Hypothesis Tests for Three-Dimensional Labelled
  Landmark Data}.
\bjournal{Scandinavian Journal of Statistics}
\bvolume{37}
\bpages{568--587}.
\end{barticle}
\endbibitem

\bibitem[\protect\citeauthoryear{Rockafellar}{1993}]{rockafellar1993lagrange}
\begin{barticle}[author]
\bauthor{\bsnm{Rockafellar},~\bfnm{R~Tyrrell}\binits{R.~T.}}
(\byear{1993}).
\btitle{Lagrange multipliers and optimality}.
\bjournal{SIAM review}
\bvolume{35}
\bpages{183--238}.
\end{barticle}
\endbibitem

\bibitem[\protect\citeauthoryear{Sch{\"a}fer and
  Strimmer}{2005}]{schafer2005shrinkage}
\begin{barticle}[author]
\bauthor{\bsnm{Sch{\"a}fer},~\bfnm{Juliane}\binits{J.}} \AND
  \bauthor{\bsnm{Strimmer},~\bfnm{Korbinian}\binits{K.}}
(\byear{2005}).
\btitle{A shrinkage approach to large-scale covariance matrix estimation and
  implications for functional genomics.}
\bjournal{Statistical applications in genetics and molecular biology}
\bvolume{4}
\bpages{Article32}.
\end{barticle}
\endbibitem

\bibitem[\protect\citeauthoryear{Shannon et~al.}{2003}]{Cytoscape03}
\begin{barticle}[author]
\bauthor{\bsnm{Shannon},~\bfnm{P.}\binits{P.}},
  \bauthor{\bsnm{Markiel},~\bfnm{A.}\binits{A.}},
  \bauthor{\bsnm{Ozier},~\bfnm{O.}\binits{O.}},
  \bauthor{\bsnm{Baliga},~\bfnm{N.~S.}\binits{N.~S.}},
  \bauthor{\bsnm{Wang},~\bfnm{J.~T.}\binits{J.~T.}},
  \bauthor{\bsnm{Ramage},~\bfnm{D.}\binits{D.}},
  \bauthor{\bsnm{Amin},~\bfnm{N.}\binits{N.}},
  \bauthor{\bsnm{Schwikowski},~\bfnm{B.}\binits{B.}} \AND
  \bauthor{\bsnm{Ideker},~\bfnm{T.}\binits{T.}}
(\byear{2003}).
\btitle{Cytoscape: a software environment for integrated models of biomolecular
  interaction networks}.
\bjournal{Genome Research}
\bvolume{13}
\bpages{2498--2504}.
\end{barticle}
\endbibitem

\bibitem[\protect\citeauthoryear{Shaw}{1990}]{10.2307/450561}
\begin{barticle}[author]
\bauthor{\bsnm{Shaw},~\bfnm{Narelle}\binits{N.}}
(\byear{1990}).
\btitle{Free Indirect Speech and Jane Austen's 1816 Revision of Northanger
  Abbey}.
\bjournal{Studies in English Literature, 1500-1900}
\bvolume{30}
\bpages{591--601}.
\end{barticle}
\endbibitem

\bibitem[\protect\citeauthoryear{{R Core Team}}{2020}]{CRAN18}
\begin{bmanual}[author]
\bauthor{\bsnm{{R Core Team}}}
(\byear{2020}).
\btitle{R: A Language and Environment for Statistical Computing}
\bpublisher{R Foundation for Statistical Computing},
\baddress{Vienna, Austria}.
\end{bmanual}
\endbibitem

\bibitem[\protect\citeauthoryear{Villani}{2009}]{Villani09}
\begin{bbook}[author]
\bauthor{\bsnm{Villani},~\bfnm{C\'edric}\binits{C.}}
(\byear{2009}).
\btitle{Optimal Transport: Old and New}.
\bpublisher{Springer}, \baddress{Berlin}.
\end{bbook}
\endbibitem

\bibitem[\protect\citeauthoryear{Ward}{1963}]{Ward63}
\begin{barticle}[author]
\bauthor{\bsnm{Ward},~\bfnm{Joe~H.}\binits{J.~H.} \bsuffix{Jr.}}
(\byear{1963}).
\btitle{Hierarchical grouping to optimize an objective function}.
\bjournal{J. Amer. Statist. Assoc.}
\bvolume{58}
\bpages{236--244}.
\bmrnumber{MR0148188 (26 \#\#5696)}
\end{barticle}
\endbibitem

\bibitem[\protect\citeauthoryear{Wilks}{1962}]{Wilks62}
\begin{bbook}[author]
\bauthor{\bsnm{Wilks},~\bfnm{S.~S.}\binits{S.~S.}}
(\byear{1962}).
\btitle{Mathematical Statistics}.
\bpublisher{Wiley}, \baddress{New York}.
\end{bbook}
\endbibitem

\end{thebibliography}

\newpage
\appendix 

\section{Most common words}\label{TAB2}

 \begin{table}[htbp]
 \begin{small}
    \begin{tabular}{|c |c| c| c| } 
\hline
& \textbf{Rank in } & \textbf{Rank in } & \textbf{Rank in} \\
\textbf{Word} & \textbf{all } & \textbf{ Dickens} & \textbf{  Austen } \\
 & \textbf{novels} & \textbf{  novels} & \textbf{novels} \\
\hline 
the & 1 & 1 & 1 \\                            
and & 2 & 2 & 3 \\                       
to & 3 & 3 & 2 \\                               
of & 4 & 4 & 4 \\                                
a & 5 & 5 & 5 \\                                
i & 6 & 6 & 7 \\                             
in & 7 & 7 & 8 \\                                
that & 8 & 8 & 13 \\                            
it & 9 & 11 & 10 \\                              
he & 10 & 10 & 16 \\                                
his & 11 & 9 & 20 \\                                 
was & 12 & 13 & 9 \\                                
you & 13 & 12 & 15 \\                                
with & 14 & 14 & 21 \\                                
her & 15 & 16 & 6 \\                             
as & 16 & 15 & 18 \\                                
had & 17 & 17 & 17 \\                                
for & 18 & 20 & 19 \\                               
at & 19 & 21 & 25 \\                              
mr & 20 & 18 & 38 \\                               
not & 21 & 26 & 12 \\                                
be & 22 & 28 & 14 \\                               
she & 23 & 31 & 11 \\                                 
said & 24 & 19 & 58 \\                                
have & 25 & 25 & 23 \\                               
\hline
\end{tabular}
\end{small}
\caption{The most common 25 words in the Austen and Dickens novels}
\end{table}

\newpage

\section{Proof for result \ref{theorem:euc mean}}\label{sec:cals for space}
Let $\lbrace \hat{\theta}_n \rbrace$ be a sequence of estimates from a sample set of graph Laplacians\\
$\lbrace \textbf{L}_1,\ldots, \textbf{L}_n\rbrace$ for a population parameter 
$\theta$. For 
$\hat\theta_n$ to be consistent it converges in probability to $\theta$ as $n \to \infty$, i.e. for any $\epsilon>0, \delta>0$ there exists a number $N$ such that for all $n\geq N$ we have $P(\vert \hat{\theta}_n-\theta\vert>\epsilon) < \delta$.

When using the power metric $d_{\alpha}$ we have the embedding space ${\mathcal M}_m$ as a Euclidean space. Hence, we know that $\hat{\eta} \in {\mathcal M}_m$ is a consistent estimator of $\eta \in 
\mathcal{M}_m$, 
as it converges in probability to $\eta$ from the law of large numbers, where $\hat\eta, \eta$ are defined in (\ref{eta1}),(\ref{eta2}). So by the continuous mapping theorem $\textbf{G}_\alpha( \hat\eta )$ converges in probability to 
$\textbf{G}_\alpha( \eta )$ as $n \to \infty$.

\def\dotMarkRightAngle[size=#1](#2,#3,#4){%
 \draw ($(#3)!#1!(#2)$) -- 
       ($($(#3)!#1!(#2)$)!#1!90:(#2)$) --
       ($(#3)!#1!(#4)$);
 %\path (#3) --node[circle,fill,inner sep=.5pt]{} ($($(#3)!#1!(#2)$)!#1!90:(#2)$);
}
\usetikzlibrary{calc}

\newcommand{\tikzAngleOfLine}{\tikz@AngleOfLine}
  \def\tikz@AngleOfLine(#1)(#2)#3{%
  \pgfmathanglebetweenpoints{%
    \pgfpointanchor{#1}{center}}{%
    \pgfpointanchor{#2}{center}}
  \pgfmathsetmacro{#3}{\pgfmathresult}%
  }

\begin{figure}[H]
\centering
  \begin{subfigure}[b]{0.45\textwidth}
 
       {
\begin{tikzpicture}
     \coordinate (A) at (3,2);%ETA 
\coordinate (B) at (4,0);%boundary
\coordinate (C) at (0,3);
\coordinate (E) at (1, 2.25); 
\coordinate (D) at ($(C)!(A)!(B)$);%mu
\coordinate (F) at (2.5,4.5);%eta hat
\coordinate (G) at ($(C)!(F)!(B)$);%mu hat
\draw (A)node[above right]{${  \textbf{G}_\alpha( \eta )  }$};
\draw (F)node[above left]{$ \textbf{G}_\alpha(\hat\eta) $};
\draw (G)node[below left]{$\hat{\mu}$};
\draw (B)node[below right]{$\mathcal{B}(\mathcal{L}_m)$}--(C)node[above left]{}--cycle;
\draw [line width=0.5mm](B)node[below right]{}--(C)node[above left]{}--cycle;
\draw (F)node[below left]{}--node[above] {$\quad\beta$}(A)node[ right]{}--cycle;
\draw (G)node[below right]{}--node[below] {$\zeta$}(D)node[above left]{}--cycle;
\draw[dashed] (A)--node[below] {} (D)node[below left]{${\mu}$};
\draw[dashed] (F)--node[below] {} (G)node[below left]{};
%\dotMarkRightAngle[size=6pt](C,A,B);
\dotMarkRightAngle[size=12pt](A,D,E);
\dotMarkRightAngle[size=12pt](B,G,F);
\end{tikzpicture}
} \caption{Case 1 }\label{fig: consistent means projection}
    \end{subfigure}
      \begin{subfigure}[b]{0.45\textwidth}
 
       {
\begin{tikzpicture}
\coordinate (A) at (4,0.75);%eta
\coordinate (B) at (2,-1);%boundary
\coordinate (C) at (0,3);%
\coordinate (E) at (1, 2.25); 
\coordinate (D) at ( 2.56,1.08);%mu
\coordinate (F) at (2,4);%eta hat
\coordinate (G) at ($(C)!(F)!(D)$);%mu hat
\coordinate (H) at (3.189189, 1.918919);%q used line.line.intersection(c(1,-1), c(2.56,1.08), c(2,4), c(4,0.5), interior.only = FALSE) function in r  library(retistruct)
\draw (A)node[above right]{${\textbf{G}_\alpha(\eta)}$};
\draw (B)node[below right]{$\mathcal{B}(\mathcal{L}_m)$};
\draw (F)node[above right]{$ \textbf{G}_\alpha(\hat\eta)$};
\draw (G)node[below left]{$\hat{\mu}$};
\draw (D)node[above]{$\vartheta$};
\draw (H)node[below ]{$q$};

\draw [ line width=0.5mm] (B)node[below left]{}-- (D)node[above left]{}--cycle;
\draw [line width=0.5mm] (C)node[below left]{}-- (E)node[above left]{}--cycle;
%\draw (B)node[below right]{}--(C)node[above left]{}--cycle;
\draw (F)node[below left]{}--node[above] {$\beta$}(A)node[above left]{}--cycle;

\draw [line width=0.5mm] (G)node[below right]{}--node[below] {$\zeta$}(D)node[above left]{}--cycle;
\draw[dashed] (A)--node[below] {} (D)node[below right]{${\mu}$};
\draw[dashed] (H)--node[below] {} (D)node[below left]{};
\draw[dashed] (F)--node[below] {} (G)node[below right]{};
\dotMarkRightAngle[size=12pt](D,G,F);
\dotMarkRightAngle[size=3pt](H,D,G);
 \tikzAngleOfLine(D)(E){\AngleStart}
 \tikzAngleOfLine(D)(A){\AngleEnd}
        \draw[] (D)+(\AngleStart:0.5cm) arc (\AngleStart:\AngleEnd-360:0.5 cm);
         \node[circle] at ($(D)+({(\AngleStart+\AngleEnd-360)/2}:0.3 cm)$) {};
%\dotMarkRightAngle[size=6pt](C,A,B);
%\dotMarkRightAngle[size=6pt](B,D,A);
\end{tikzpicture}
} \caption{Case 2 }\label{fig: consistent means projection case2}
    \end{subfigure}
    \caption{2D representations of the possibly high dimensional faces of $\mathcal{L}_m$ illustrating convergence of means.}
\end{figure}
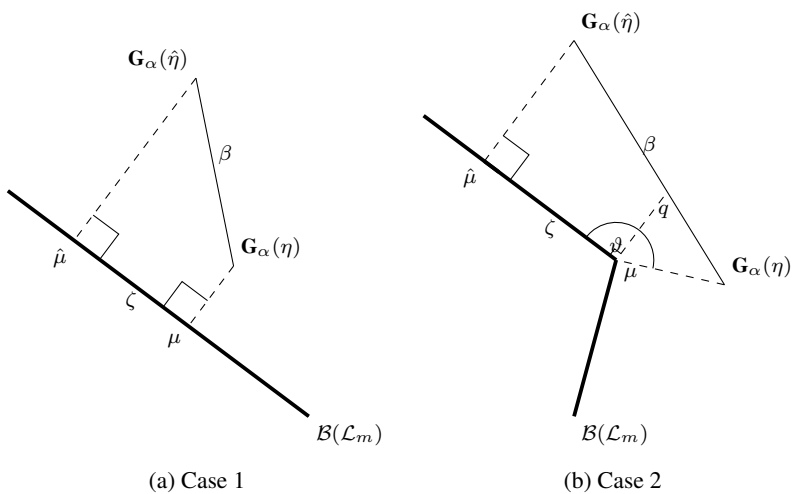

 We now need to show the convergence in probability holds when we project to $\mathcal{L}_m$. Recall that $\mathcal{L}_m\subset\mathcal{M}_m$ and both spaces have dimension $\frac{m(m-1)}{2}$. 
 { \citet{ginestet2017hypothesis} showed that $\mathcal{L}_m$ is a closed compact convex subset of an affine 
space, i.e. each face of $\mathcal{L}_m$ is isometric to a subset of $[0,\infty)^k\times \mathbb{R}^{d-k}$ for some $k>0$. Hence the interior of each 
face of 
$\mathcal{L}_m$ has zero curvature, and we denote the boundary of $\mathcal{L}_m$ as $\mathcal{B}(\mathcal{L}_m)$.} Let $\beta = \vert \textbf{G}_\alpha( \hat\eta ) - \textbf{G}_\alpha( \eta ) \vert $ and $\zeta = \vert\hat{\mu}-\mu\vert$.

 There are three cases to consider:

 \begin{itemize}
\item Case 1: ${\mu}$ is in $\mathcal{B}(\mathcal{L}_m)$ but not on a corner. In this case the estimator behaves as in Figure \ref{fig: consistent means projection}.
The estimator $ \textbf{G}_\alpha( \hat\eta ) $ is orthogonally projected to $\hat{\mu}$, hence due to Pythagoras' theorem it is clear $\zeta\leq\beta$.
\item Case 2:  $\mu$ is on a corner of $\mathcal{B}(\mathcal{L}_m)$.  In this case the estimator behaves as in Figure \ref{fig: consistent means projection case2}. Clearly $\frac{\pi}{2}\leq \vartheta\leq\pi$ as $\mathcal{L}_m$ is convex \citep{ginestet2017hypothesis}. We consider a point $q$ along the line between $\textbf{G}_\alpha( \hat\eta )$ and $\textbf{G}_\alpha( \eta ) $ such that the angle between $\hat{\mu}$, $\mu$ and $q$ is $\frac{\pi}{2}$. Note $\zeta \le | \textbf{G}_\alpha( \hat\eta ) - q |$ following identical arguments as in case 1, and clearly $|\textbf{G}_\alpha( \hat\eta )  - q | \le \beta$. 
Hence $\zeta\leq\beta$.
\item Case 3: ${\mu}$ is in the interior of $\mathcal{L}_m$, and so $\textbf{G}_\alpha( \eta ) = \mu$ and by Pythagoras' theorem $\zeta \leq \beta$. 
 \end{itemize}

From the consistency of $\textbf{G}_\alpha( \hat\eta )$ for any $\epsilon>0, \delta>0$  
there exists an $N$ such that for $n\geq N$ then $P(\vert  \textbf{G}_\alpha( \hat\eta ) - \textbf{G}_\alpha( \eta )   \vert>\epsilon)<\delta$. 
So, in all three cases we deduce that 
$$P( \vert \hat\mu - \mu \vert > \epsilon ) = P( \zeta > \epsilon )  \le P( \beta > \epsilon) =  P(\vert  \textbf{G}_\alpha( \hat\eta ) - \textbf{G}_\alpha( \eta )   \vert>\epsilon)<\delta$$
and so $\hat\mu$ converges in probability to $\mu$, i.e. $\hat\mu$ is a consistent estimator for $\mu$.

\newpage 

\section{Means for Austen's and Dickens' novels}\label{sec:other means}
\begin{figure}[htbp]
\includegraphics[width=6cm]{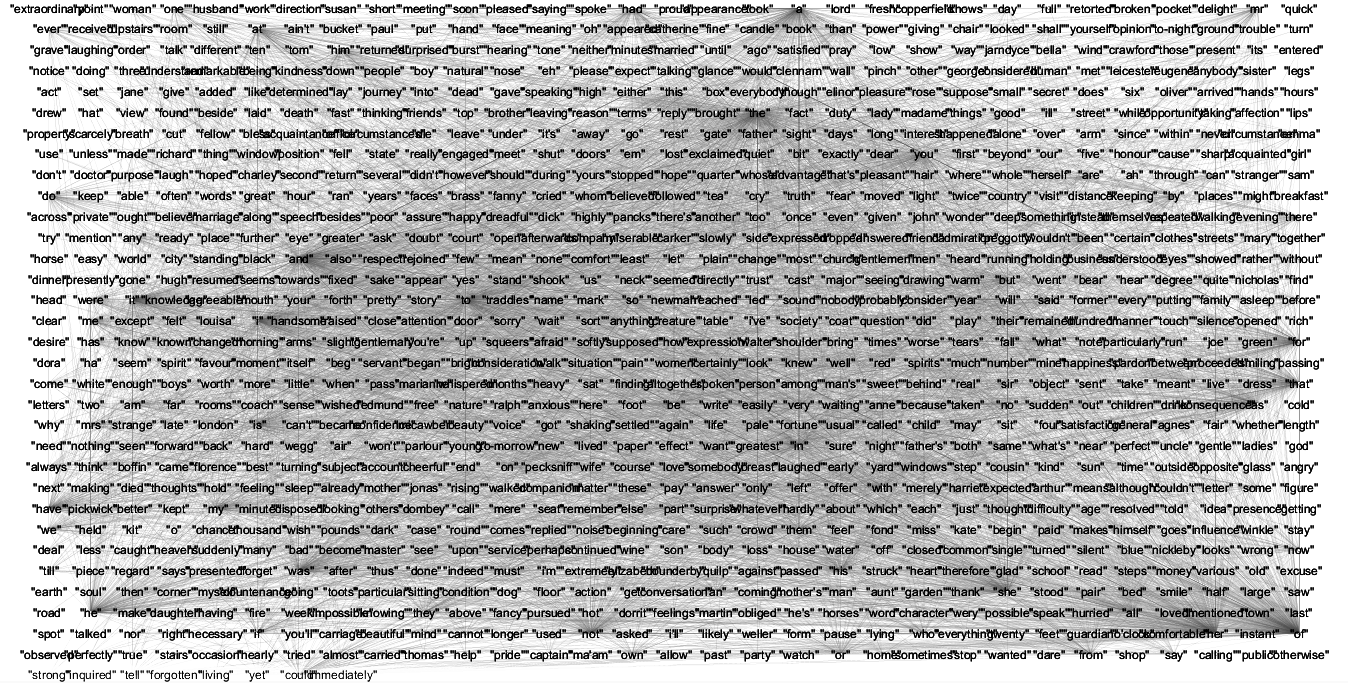}
\includegraphics[width=6cm]{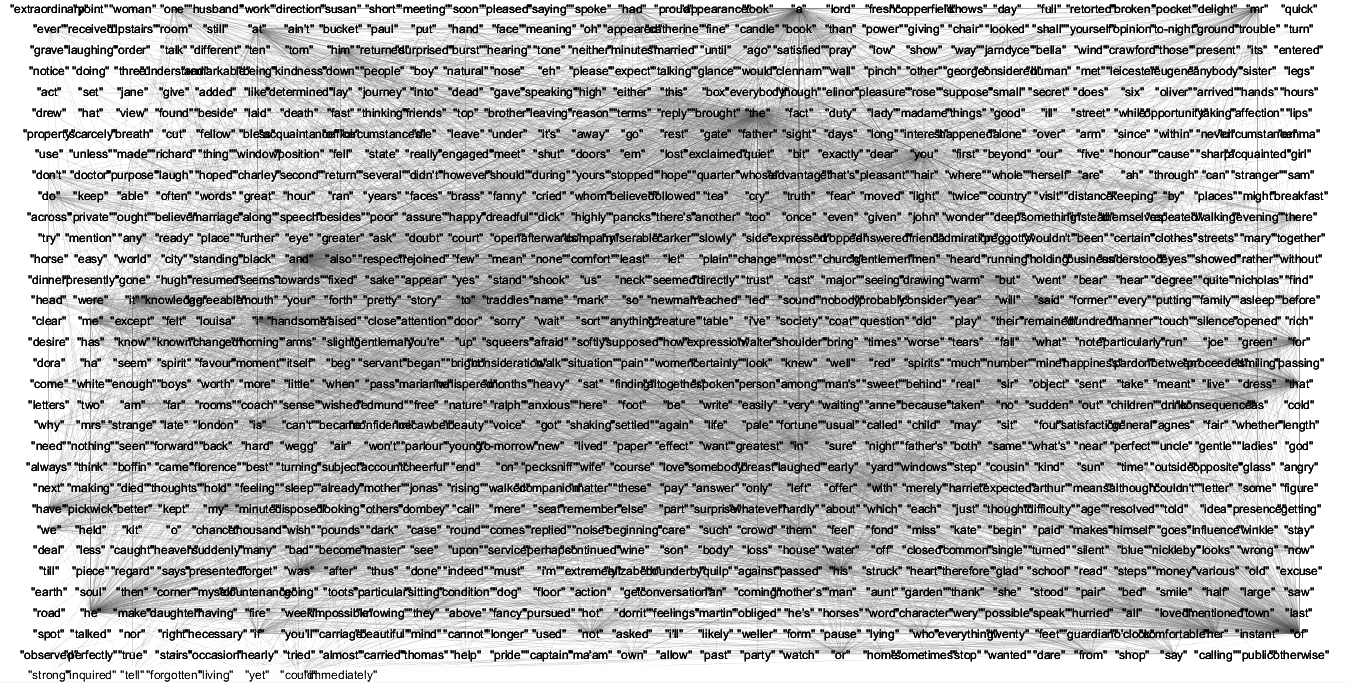}\\
\includegraphics[width=6cm]{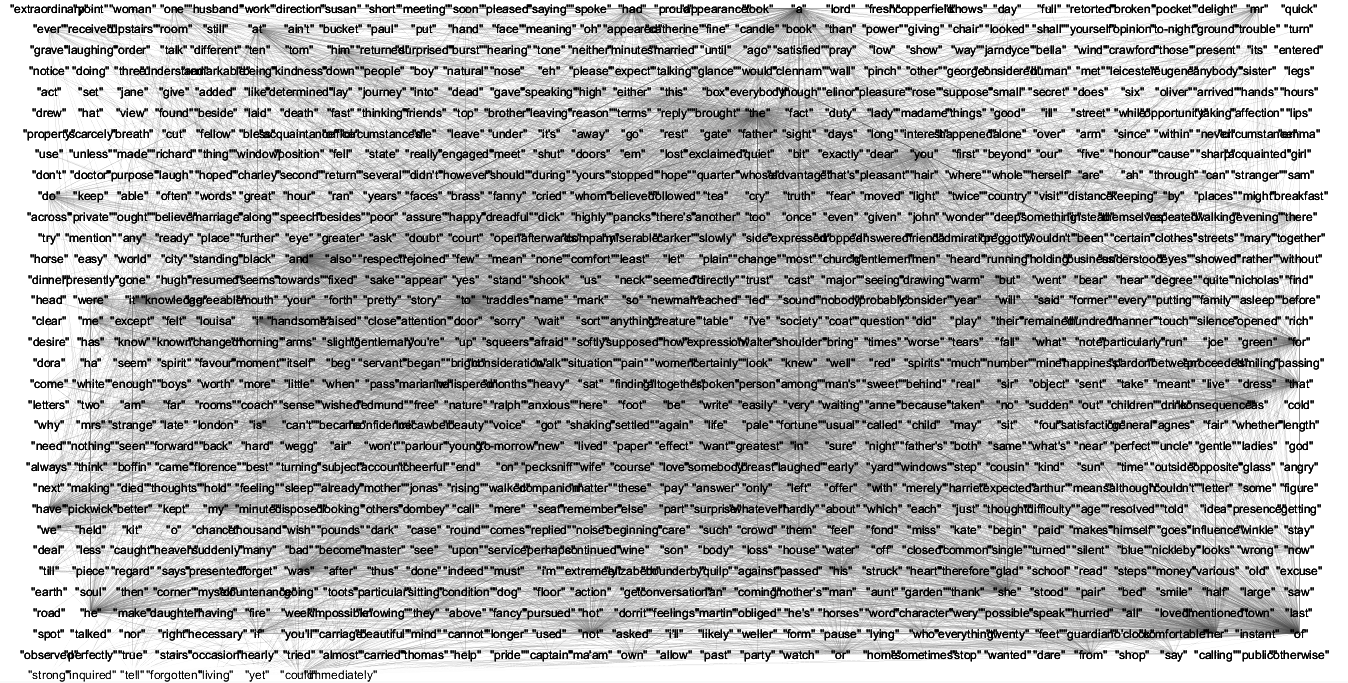}
\includegraphics[width=6cm]{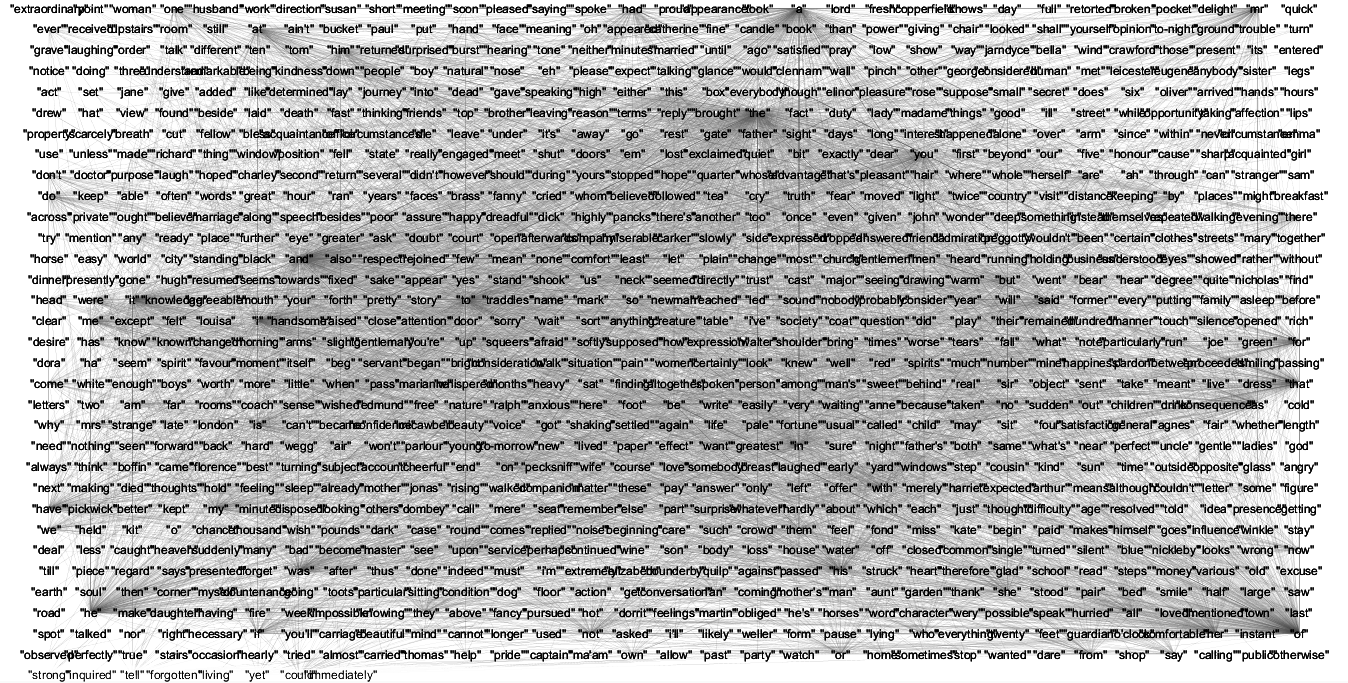}\\
\includegraphics[width=6cm]{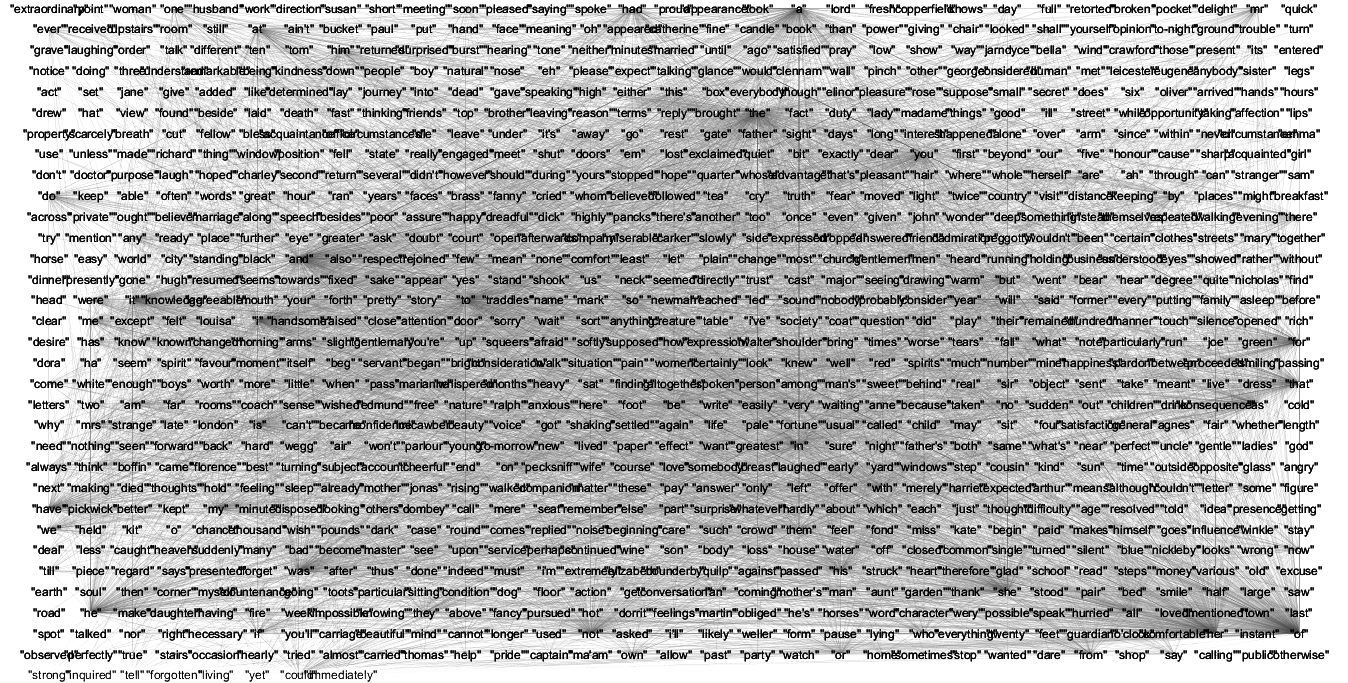}
\includegraphics[width=6cm]{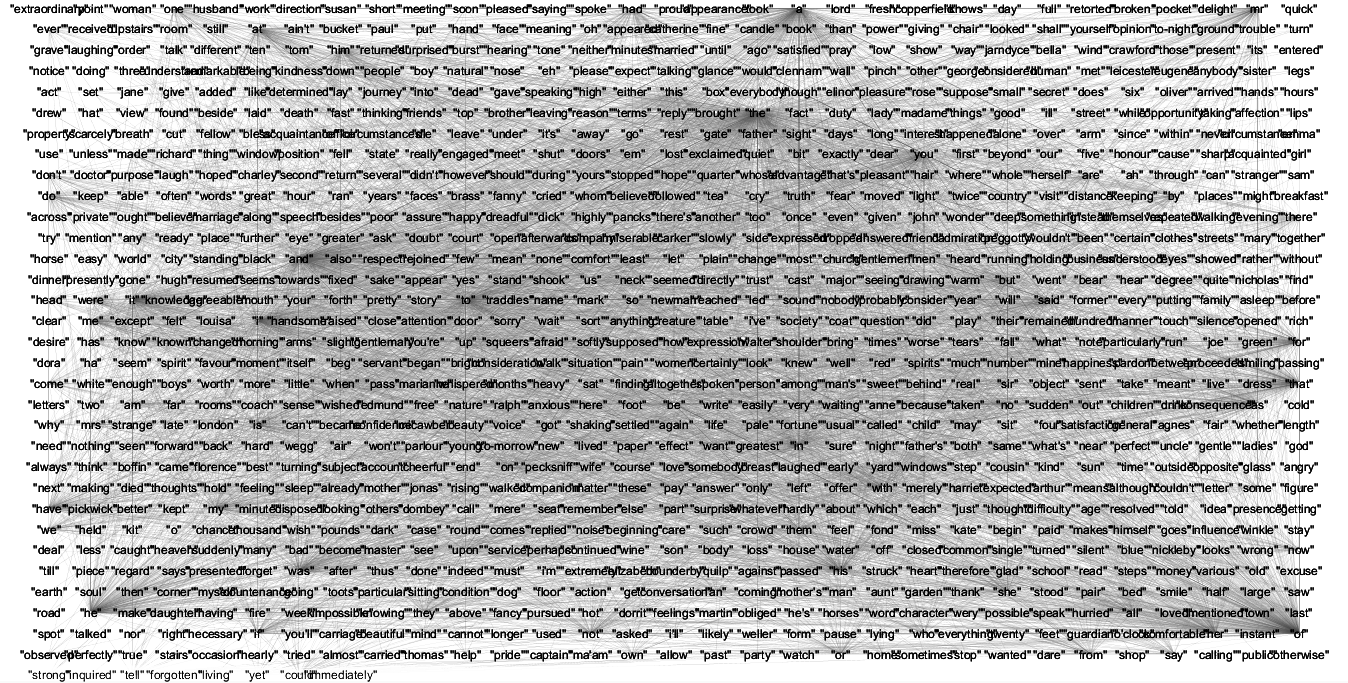}
\caption{ \footnotesize{\textit{The means of (left) Austen's novels and (right) Dickens' novels using $d_1$ (first row), $d_{\frac{1}{2}}$ (middle row) 
and $d_{\frac{1}{2}, S}$ (bottom row) based on the top m=1000 word pairs. Zoom in for more detail. }}}\label{hugeplots proc}        \end{figure}

\end{document}